# Approximate Bayesian inference of seismic velocity and pore pressure uncertainty with basin modeling, rock physics and imaging constraints

Anshuman Pradhan[1*], Huy Q. Le[2], Nader C. Dutta[2], Biondo Biondi[2] & Tapan Mukerji[1]

[1]Department of Energy Resources Engineering, [2]Department of Geophysics

Stanford University

[*]pradhan1@stanford.edu

## ABSTRACT

We present a methodology for quantifying seismic velocity and pore pressure uncertainty that incorporates information regarding the geological history of a basin, rock physics, well log, drilling and seismic data. In particular, our approach relies on linking velocity models to the basin modeling outputs of porosity, mineral volume fractions and pore pressure through rock physics models. We account for geological uncertainty by defining prior probability distributions on uncertain parameters and performing Monte Carlo basin simulations. We perform probabilistic calibration of the basin model outputs by defining data likelihood distributions to represent well data uncertainty. Rock physics modeling transforms the basin modeling outputs to give us multiple velocity realizations used to perform multiple depth migrations. We present an approximate Bayesian inference framework which uses migration velocity analysis in conjunction with well data for updating velocity and basin modeling uncertainty. We apply our methodology in 2D to a real field case from Gulf of Mexico and demonstrate that our methodology allows building a geologic and physical model space for velocity and pore pressure prediction with reduced uncertainty.

## INTRODUCTION

The goal of most seismic velocity inverse problems is to obtain models which match the kinematics and/or amplitudes of the reflections observed in seismic data. Velocity inverse problems are ill-posed, and it is generally the case that several models equivalently honor the seismic data constraints. Numerous strategies have been researched and developed for efficiently solving these inverse problems (Biondi, 2006). However, a key research challenge that still exists is ensuring the inverted models 1) honor geologic plausibility and variability and 2) are physically valid. By geologic plausibility and variability, we refer to the agreement of the velocity model space with prior geologic knowledge about geo-history (events and processes that occurred in the geologic past). This is a crucial problem because present-day rock velocities are directly coupled to various processes like deposition, compaction, and fluid flow that the subsurface was exposed to in the geological past. By physically validity, we refer to agreement of the velocity model with other physical quantities such as pore pressure, time to depth relations of seismic horizons, lithology and drilling indicators such as fracture pressure and overburden stress. This is a crucial problem because, in most cases, such physical constraints are not explicitly incorporated into objective functions for velocity inversion. In this paper, we advocate that the goal of seismic velocity estimation should be to estimate rock velocities that satisfy these geological and physical constraints in addition to seismic data and imaging constraints. A velocity estimation method which yields rock velocities will significantly boost the reliability of depth conversion of earth models and render velocity models useful in other industry applications such as reservoir characterization and reliable estimation of pore pressure, fracture or overburden stresses. In this paper, we distinguish between seismic velocity and rock elastic velocity. While seismic velocity is affected by how the seismic data are acquired and processed, rock velocity is an intrinsic property of the rock being sampled and is determined mainly by its composition, their elastic properties, pore microgeometry, and alteration of constituents and microgeometry during burial processes.

Incorporating geological and physical constraints in velocity inversion is non-trivial to address. Difficulties primarily arise because relevant information can be available in diverse forms ranging from abstract notions about paleo geological events which potentially affected the present-day velocity distribution to different types of well measurements like mud-weight, porosity or temperature, which are implicitly linked to velocity, as well as structural interpretations of geologic horizons. The challenge is to quantitatively integrate all these diverse types of information into the inversion procedure. A common procedure is to incorporate a regularization term in the objective function of the inverse problem such that it penalizes solution models which deviate from an established prior geological model. Clapp et al. (2004) propose the use of steering filters as regularization operators which encourage solutions to replicate the structure of an interpreted model of the subsurface. While such regularization strategy generates velocity models that conform to the present-day geologic structure, a limitation is that it does not guarantee that the ensuing models are physical rock velocities. Depending on the depositional, thermal and pressure geohistory, various mechanical and chemical sediment compaction mechanisms such as compaction disequilibrium, smectite-illite transition (S-I), and quartz cementation could have been at play in a basin. These effects distinctively affect the present-day rock composition and texture, and in turn, the rock velocities.

Several authors have addressed the problem of incorporating geological and physical constraints in velocity inversion by employing basin modeling and rock physics (Petmecky et al., 2009; Brevik et al., 2011; Bachrach et al., 2013; Dutta et al., 2014; Li, 2014; De Prisco et al., 2015). Basin modeling presents a quantitative framework for simulating geologic processes through time and yields various earth properties such as porosity, pore-pressure, effective stress and mineral volume fractions (Hantschel and Kauerauf, 2009). Rock physics provides physical models of the rock microstructure, which can be used to relate basin modeling generated properties to elastic properties, and consequently to velocities (Figure 1). Brevik et al. (2011, 2014), De Prisco et al. (2015) and Szydlik et al. (2015) lay out a framework for integrating basin modeling with geophysics in what has been termed as Geophysical Basin Modeling. Their proposed workflow entails modeling the basin in the area of interest and subsequently employing rock physics based models to generate the initial velocity model and model parameter covariance to be used in the velocity inversion process. A key contribution of their approach is that it facilitates constraining the prior space of velocity model parameters to be geologically and physically consistent.

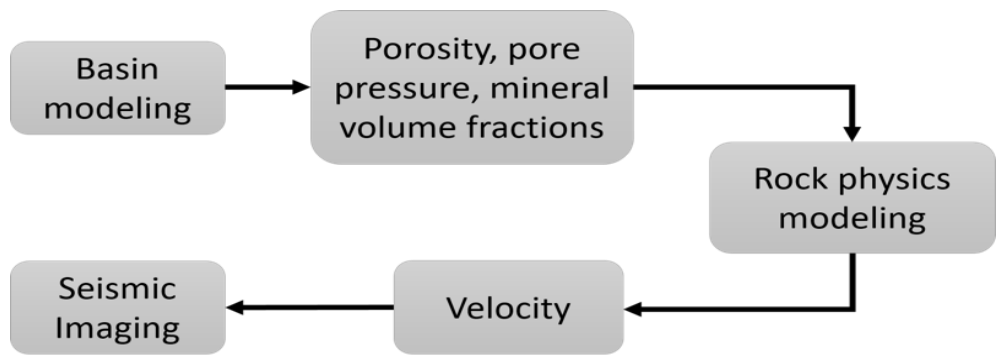

Figure 1: The integrated workflow used in this paper.

Even though basin modeling presents a quantitative framework for modeling of prior geologic beliefs, a major limitation associated with the method is the high degree of uncertainty associated with the basin modeling input parameters such as the lithologic compaction functions, lithology-specific porosity-permeability relations, present-day structural and stratigraphic model and boundary conditions such as paleo heat flow. The conventional approach to specifying these parameters is to perform a deterministic calibration of modeling outputs to well data. However, since our final goal is to generate velocity models which match both well and seismic data, it is necessary to rigorously quantify and propagate any basin modeling uncertainties into the velocity model space. De Prisco et al. (2015) identify these challenges and propagate basin modeling uncertainties into velocity models by considering certain geologic scenarios and selecting one which exhibits best match to well data. Subsequently, several input parameter values within this scenario are perturbed to generate multiple basin modeling outputs. One crucial aspect is that, in many cases, the calibration data at the wells are also uncertain. Thus, any uncertain geologic scenarios or parameters included in the analysis should be considered in accordance with the probabilistic model for the data uncertainty.

In this paper, we propose an integrated Bayesian framework for velocity modeling with geohistory and rock physics constraints, which facilitates rigorous propagation of basin modeling uncertainties into the velocity prior models. Our approach entails specifying prior probability distributions on uncertain basin modeling input parameters and likelihood models for well calibration data. We perform multiple basin modeling runs using realizations of the uncertain parameters obtained through Monte Carlo sampling of the prior uncertainties. Posterior basin modeling realizations are then selected according to the likelihood computed from the well calibration data. These accepted posterior realizations are subsequently linked to velocity through a calibrated friable shaly-sand rock physics model. These velocity models can then serve as geologically consistent and physically valid priors for the seismic velocity inversion problem.

We also demonstrate how our proposed Bayesian inference framework can be leveraged to go beyond velocity and quantify uncertainties of various other earth properties, specifically the basin modeling outputs such as pore-pressure, porosity or smectite content of host rocks (mostly shale in our study area). These properties act as essential variables in common industry tasks such as pore pressure prediction and reservoir characterization. The output properties from basin models typically do not find explicit usage in these applications since basin models lack any spatial calibration away from the wells. Rather, velocity models obtained by inversion are widely used, both explicitly and implicitly, in the estimation of these properties. For instance, velocity models are used as inputs in rock physics transforms to generate pore pressure volumes (Dutta, 2002). Velocity models are also used as low-frequency constraints in the inversion of elastic impedances or reservoir properties. This is so because inverted velocity models match the kinematics of the seismic data and thus provide crucial spatial data constraints. However, as emphasized earlier, inversion frameworks do not explicitly stipulate that the velocity model be geological or physical. This necessitates an additional modeling step to transform seismic velocities to rock velocities (Al-Chalabi, 1994; Dutta et al., 2014). A limitation with this approach is that it precludes consistent propagation of uncertainty from geology to seismic. Note that in our proposed methodology we parameterize velocity through basin modeling outputs in an integrated framework, thus rendering feasible the consistent propagation of uncertainties from geology to seismic velocity. This integrated formulation also presents opportunities to use seismic data to inform the geologic uncertainty space. To accomplish this, we propose an approximate Bayesian inference scheme employing migration velocity analysis which facilitates uncertainty quantification of velocity and earth properties such as pore-pressure, porosity or smectite content consistently with seismic data kinematics.

In summary, the intended contributions of the paper are:

1) Formulation of a Bayesian framework for velocity model building by Geophysical Basin Modeling. This framework facilitates rigorous propagation of prior geologic and well data uncertainty to generate a geologically consistent and physically valid model space for traditional velocity inversion techniques

2) Proposal of a method for performing approximate inference of velocity from seismic data using migration velocity analysis. Within our integrated framework, this method allows constraining basin models with seismic data, thus generating a seismically informed uncertainty space for earth properties like pore-pressure or porosity.

The paper is organized into three sections: 1) 'Methodology', 2) 'Real Case application' and 3) 'Discussion and Conclusions'. We begin the paper by providing a detailed treatment of the theoretical nuances of our workflow and the methods we employ. Subsequently, we discuss application of our methodology to a 2D basin in the Northern Gulf of Mexico. With the help of this example, we demonstrate that our integrated Bayesian framework generates multiple velocity models which agree with the seismic data kinematics and especially honor the geologic uncertainty and rock physical bounds. We also demonstrate an application of migration velocity analysis constrained pore pressure prediction. We show how our method was effective for reducing prior geologic uncertainty on velocity and pore pressure. Finally, we discuss advantages and limitations of the presented work, along with future research directions.

METHODOLOGY

We incorporate geologic and physical constraints into our analysis by linking velocity models to basin modeling outputs through rock physics models. In basin modeling, given a set of input parameters $\boldsymbol{b}$, differential equations are solved through geologic times for simulating the effects of desired processes such as sedimentation, compaction, fluid flow and heat flow (Hantschel and Kauerauf, 2009). The outputs from the basin model, $\boldsymbol{h}$, include present-day distributions of various earth properties such as porosity, pore-pressure, temperature and mineral volume fractions, and can be represented as

$$\boldsymbol{h} = g_{BM}(\boldsymbol{b}). \qquad (1)$$

Here, $g_{BM}(.)$ is the functional notation for basin modeling simulation equations. The link with present-day velocity $\boldsymbol{v}$ is established through the rock physics model:

$$\boldsymbol{v} = g_{RPM}(\boldsymbol{h}). \qquad (2)$$

Given the above parameterization, our goals are to obtain basin and velocity models which are consistent with any prior beliefs on geological uncertainty while matching the observed basin modeling calibration data $\boldsymbol{d}_{w,obs}$ and seismic data $\boldsymbol{d}_{s,obs}$. Note that basin modeling outputs $\boldsymbol{h}$ are obtained across the entire earth model but the calibration is performed only at discrete well locations. Spatial constraints away from the wells will be imposed by evaluating whether the corresponding velocity model $\boldsymbol{v}$ is in agreement with $\boldsymbol{d}_{s,obs}$. A standard approach to establishing this involves forward modeling kinematics of seismic reflections

$$\boldsymbol{k}_m = g_{WP}(\boldsymbol{v}), \qquad (3)$$

and matching to observed kinematics. Here, $g_{WP}(.)$ is the wave-propagation model. Details on extraction of kinematic information $\boldsymbol{k}_m$ is provided in a later sub-section.

Bayes' rule provides the solution to the joint inference problem stated above:

$$f(\boldsymbol{b}, \boldsymbol{v}|\boldsymbol{d}_{w,obs}, \boldsymbol{d}_{s,obs}) = \frac{f(\boldsymbol{d}_{w,obs}, \boldsymbol{d}_{s,obs}|\boldsymbol{b}, \boldsymbol{v})f(\boldsymbol{b}, \boldsymbol{v})}{\int_{\boldsymbol{b}, \boldsymbol{v}} f(\boldsymbol{d}_{w,obs}, \boldsymbol{d}_{s,obs}|\boldsymbol{b}, \boldsymbol{v})f(\boldsymbol{b}, \boldsymbol{v}) \, d\boldsymbol{b} \, d\boldsymbol{v}}. \qquad (4)$$

Here, $f(\boldsymbol{b}, \boldsymbol{v}|\boldsymbol{d}_{w,obs}, \boldsymbol{d}_{s,obs})$ is the posterior distribution, $f(\boldsymbol{d}_{w,obs}, \boldsymbol{d}_{s,obs}|\boldsymbol{b}, \boldsymbol{v})$ is the data likelihood distribution quantifying the conditioning to the two data types considered and $f(\boldsymbol{b}, \boldsymbol{v})$ is the prior joint distribution of basin model input parameters and velocity. In theory, it is possible to sample the exact posterior by generating samples $\{\boldsymbol{b}, \boldsymbol{v}\}$ of prior $f(\boldsymbol{b}, \boldsymbol{v})$, forward modeling the data and accepting or rejecting the prior samples according to their data likelihood. However, the high dimensionality of the joint model and joint data space will typically necessitate evaluation of a large number of prior samples. The computational demands are further exacerbated by the fact that a single prior sample evaluation requires three expensive forward modeling runs (equations [1], [2] and [3]). This will significantly limit the practicality of the approach. In order to improve the computational efficiency, we propose making the following approximations to the joint posterior distribution

$$f(\boldsymbol{b}, \boldsymbol{v}|\boldsymbol{d}_{w,obs}, \boldsymbol{d}_{s,obs}) = c_0 f(\boldsymbol{d}_{w,obs}, \boldsymbol{d}_{s,obs}|\boldsymbol{b}, \boldsymbol{v})f(\boldsymbol{b}, \boldsymbol{v})$$

$$\approx c_0 f(\boldsymbol{d}_{s,obs}|\boldsymbol{b}, \boldsymbol{v})f(\boldsymbol{d}_{w,obs}|\boldsymbol{b}, \boldsymbol{v})f(\boldsymbol{v}|\boldsymbol{b})f(\boldsymbol{b})$$

$$\approx c_0 f(\boldsymbol{d}_{s,obs}|\boldsymbol{v})f(\boldsymbol{v}|\boldsymbol{b})f(\boldsymbol{d}_{w,obs}|\boldsymbol{b})f(\boldsymbol{b}). \qquad (5)$$

Here, $c_0$ denotes the normalization constant specified in the denominator of equation [4]. In the second line of equation [5], we decompose the joint likelihood by assuming that basin modeling data $\boldsymbol{d}_{w,obs}$ and seismic data $\boldsymbol{d}_{s,obs}$ are conditionally independent given $\boldsymbol{b}$ and $\boldsymbol{v}$. In the third line, we assume that $\boldsymbol{d}_{w,obs}$ is independent of $\boldsymbol{v}$ given $\boldsymbol{b}$ and vice versa for $\boldsymbol{d}_{s,obs}$. Note that by making the above approximations, we gain several computational conveniences:

1) The joint distribution is factored into multiple lower dimensional distributions, each of which will be easier to model and sample from.

2) By assuming the first approximation in equation 5, we dissociated the well and seismic data likelihoods. Finding models conditioned to $\boldsymbol{d}_{w,obs}$ will be significantly easier than honoring both $\boldsymbol{d}_{s,obs}$ and $\boldsymbol{d}_{w,obs}$ given that the dimensionality of $\boldsymbol{d}_{w,obs}$ is much lower than that of $\boldsymbol{d}_{s,obs}$. This observation is complemented by the fact that the dimensionality of $\boldsymbol{b}$, in many cases, will be lower than $\boldsymbol{v}$. Basin models operate at coarser scales than seismic velocities. The vector $\boldsymbol{b}$ typically consists of global parameters of the lithological layers of the earth model and its boundary conditions. The vector $\boldsymbol{v}$, on the other hand, is defined on a finer-scale grid of the earth.

3) By assuming the second approximation in equation 5, we dissociated the basin modeling and seismic imaging components. Consequently, during conditioning to $\boldsymbol{d}_{w,obs}$, only forward model $g_{BM}(.)$ needs to be evaluated. Forward models $g_{RPM}(.)$ and $g_{WP}(.)$ will now be evaluated on samples from the uncertainty space informed by $\boldsymbol{d}_{w,obs}$, i.e., $f(\boldsymbol{d}_{w,obs}|\boldsymbol{b})f(\boldsymbol{b})$. In the original joint formulation of the posterior, these forward models need to be evaluated on the unconditional prior realizations.

Even after factoring the joint distribution, efficient sampling from the seismic likelihood distribution might still be a concern given the high dimensionality of $\boldsymbol{d}_{s,obs}$ and $\boldsymbol{v}$. In order to lend further efficiency to our approach, we will employ approximate Bayesian computation (ABC), as described in detail later in this section. We will now present a detailed discussion of the main components of our proposed methodology, which are: 1) modeling of prior geological uncertainty, 2) Monte-Carlo basin modeling and probabilistic calibration, 3) rock physics modeling and migration velocity analysis, 4) velocity uncertainty quantification using approximate Bayesian inference & importance sampling and 5) Pore pressure prediction and uncertainty analysis.

**Modeling of prior geological uncertainty**

We specify prior geological uncertainty in a basin as the probability distribution $f(\boldsymbol{b})$. Modeling of $f(\boldsymbol{b})$ is a highly subjective exercise and will vary from basin to basin. Several sources of uncertainty can be considered as listed below.

1. Deposition related geological processes such as deposition and erosion of sediments, salt movement.

2. Compaction, diagenesis and overpressure related processes such as mechanical compaction, smectite-illite transformation, quartz cementation and hydrocarbon generation and migration

3. Lithological models and properties such as porosity-depth trends, permeability-porosity model and thermal properties

4. Fluid flow and heat-flow boundary conditions such as paleo heat flow, paleo water depth and sediment-water interface temperature

5. Structural interpretations of geological horizons and faults

The composition and dimensionality of the vector $\boldsymbol{b}$ will depend on which uncertain processes and variables are chosen to be modeled. While it would be ideal to incorporate all uncertainty sources that could potentially affect the present-day velocities, high dimensionality of $\boldsymbol{b}$ will limit the efficiency of the approach. Another practical aspect to consider is the ease and feasibility of formulating a probabilistic model for any uncertain variable in question. Prior uncertainty for some variables will be simple to model. For instance, uncertainty regarding the occurrence of any particular diagenetic process can be modeled as a Bernoulli distribution. Many lithological model parameters are single dimensional continuous variables which can be modeled by standard probability distributions such as the Gaussian distribution. In other cases, prior uncertainty will not be straightforward to quantify and might require advanced modeling strategies. This might be the case if spatial uncertainty is associated with boundary conditions input to the basin model or structural uncertainty associated with interpreted horizons or faults. Methodologies developed in the geostatistical literature (Caers, 2011) are well-suited to handle such spatial uncertainties.

**Monte-Carlo basin modeling and probabilistic calibration**

Once the prior is specified, it is sampled by Monte-Carlo sampling to generate multiple possible geological scenarios. For each given sample, a basin simulation through geological times is executed to obtain present-day models of various earth properties $\boldsymbol{h}$ (equation 1). The subsequent task is to condition the prior models based on the calibration data $\boldsymbol{d}_{w,obs}$. To this end, we employ the likelihood probability distribution $f\left(\boldsymbol{d}_{w,obs} \mid \boldsymbol{b}\right)$ to quantify mismatch between $\boldsymbol{d}_{w,obs}$ and corresponding elements of $\boldsymbol{h}$ at well locations, denoted by $\boldsymbol{d}_w$. $\boldsymbol{d}_{w,obs}$ typically consists of different data types such as porosity, mudweight and temperature data. Thus, $\boldsymbol{d}_{w,obs}$ is given as

$$\boldsymbol{d}_{w,obs} = [\boldsymbol{d}_{w,obs}^1, \boldsymbol{d}_{w,obs}^2, \ldots, \boldsymbol{d}_{w,obs}^n]^T, \qquad (6)$$

where, $\boldsymbol{d}_{w,obs}^i$ is the $i^{th}$ calibration data set, with $i = \{1, \ldots n\}$. By assuming conditional independence of each data type from the others given $\boldsymbol{b}$, the likelihood distribution can be expressed as

$$f(\boldsymbol{d}_{w,obs} \mid \boldsymbol{b}) = \prod_{i=1}^{n} f(\boldsymbol{d}_{w,obs}^i \mid \boldsymbol{b}) \qquad (7)$$

The above decomposition is particularly useful since each data type might impose differing constraints on the prior models. For instance, mudweight data provides upper bounds on pore pressure components of $\boldsymbol{d}_w$ while temperature data are direct measurements. Consequently, it will be necessary to employ different likelihood distributions for each data type. The likelihood distribution for $i^{th}$ data type can then be defined as

$$f(\boldsymbol{d}_{w,obs}^i \mid \boldsymbol{b}) = f(\frac{\|\boldsymbol{d}_w^i - \boldsymbol{d}_{w,obs}^i\|}{\sigma}) \qquad (8)$$

Here, $\|.\|$ is a distance measure and $\boldsymbol{d}_w^i$ are the elements of $\boldsymbol{d}_w$ corresponding to same data-type as $\boldsymbol{d}_{w,obs}^i$. $\boldsymbol{\sigma}$ is used to control the degree of accuracy desired in the fitting to $\boldsymbol{d}_{w,obs}$ and can be assigned according to the uncertainty present in the data. The likelihood values of all the prior basin models are employed during selection of posterior models which honor observed well data.

**Rock physics modeling and migration velocity analysis**

The next component of our workflow is to go from basin modeling outputs to predictions of rock velocities, i.e., sampling $f(\boldsymbol{v}|\boldsymbol{b})$. The basin modeling outputs used to predict velocities include the porosity, mineral composition (which depends on the amount of diagenesis and cementation), the lithostatic pressure and the pore pressure. It is necessary to employ a quantitative model of the rock microstructure which allows calculation of the effective elastic moduli of the rock, given the composition of the rock constituents, porosity, and the effective stress at which the rock exists. Several models are available in the literature (Mavko et al., 2009) and an appropriate model should be chosen depending on the requirements of the problem and calibrated to the rock elastic properties as observed in the well logs. For the case study presented in this paper, we employ the constant-clay model for shaly sands (Avseth et al., 2005) to model isotropic velocities using basin modeling outputs. In this model, the effective elastic properties of the rock at any rock composition, as specified by the basin model outputs, is derived by interpolating the elastic properties of the well-sorted grain pack (at the high-porosity end) and the effective mineral properties (at the zero porosity end) using the modified lower Hashin-Shtrikman bound (Mavko et al., 2009). The well-sorted grain pack is modeled as an elastic sphere pack subject to the in-situ effective stress. The rock effective stresses are calculated by Terzaghi's principle with the basin modeling outputs of pore pressure and lithostatic pressure. If the goal is to generate a prior model space for anisotropic velocities, $g_{RPM}(.)$ in equation [2] should be chosen to be an appropriate anisotropic rock model. Bandyopadhyay (2009) discusses several anisotropic rock models for shales and sands.

These velocity models serve as a prior for the velocity inference part. The subsequent step is to condition the prior velocity realizations with the kinematics of the observed seismic reflections. Kinematic information can be extracted either in the seismic data domain or the seismic image domain. We use the image domain since performing the analysis in image domain has several advantages over data domain as discussed by Biondi (2006). Prestack seismic data are imaged by performing depth migration with each prior velocity realization. Velocity models consistent with the data kinematics will generate well-focused images. Focusing quality of the imaged reflectors are evaluated using angle-domain common image gathers (ADCIGs; Sava and Fomel, 2003; Biondi and Shan, 2002). ADCIGs describe variability of seismic events across different reflection angles at any particular subsurface location. Optimal focusing of the reflections corresponds to flat events across the reflection angle axis. Flatness of the events in the ADCIG can thus be employed as a criterion for conditioning the prior velocity realizations to the observed seismic data.

Residual moveout analysis (RMO; Al-Yahya, 1989) is a commonly employed method for quantifying the deviation from flatness of the ADCIG events. We parameterize the RMO function in the angle domain by the single parameter $\rho$ as described by Biondi (2006)

$$\rho(x, y, z) = {v_t(x, y, z)} \big/ {v_\rho(x, y, z)}. \tag{9}$$

Here, $x$, $y$ and $z$ represent the coordinates of the earth location. $v_t(x, y, z)$ is the true migration velocity at given subsurface location, i.e., the velocity that would flatten the ADCIGs. $v_\rho(x, y, z)$ is the velocity being evaluated, i.e., the prior velocity in our case. For each prior velocity model, RMO analysis is performed. RMO analysis consists of scanning through possible $\rho$ values, applying residual moveout correction to the ADCIGs using the RMO function and picking the $\rho$ value $\rho_{maxS}(x,y,z)$ which maximizes the semblance of the ADCIGs. Measure of how far the prior velocity is from the optimally focusing velocity is given by the RMO error

$$\epsilon(x,y,z) = \rho_{maxS}(x,y,z) - 1. \tag{10}$$

Deterministic migration velocity analysis techniques typically seek to perform local optimization of an initial estimate of the velocity model. Specifically, the local errors $\epsilon(x,y,z)$ are translated into local updates or perturbations of the initial velocity model according to inverse of the wave propagation forward model linking $v(x,y,z)$ to the $\epsilon(x,y,z)$. Note that in our case the velocity model is implicitly parameterized by basin modeling parameters $\boldsymbol{b}$. These parameters are global in the sense that these parameters are not defined on the earth model grid, rather, they control global variability of the model. Correspondingly, our goal is to find models which satisfy a global measure of conformance to kinematic constraints. For this purpose, we employ the mean of absolute RMO errors across all grid locations

$$\epsilon_{global} = \frac{\sum_x \sum_y \sum_z |\epsilon(x,y,z)|}{\sum_x \sum_y \sum_z 1}. \tag{11}$$

**Velocity uncertainty quantification: Approximate Bayesian inference & importance sampling**

In this sub-section, we discuss approximations and strategies employed to generate velocity models conditioned to both $\boldsymbol{d}_{w,obs}$ and $\boldsymbol{d}_{s,obs}$. The posterior distribution we aim to sample is

$$f(\boldsymbol{b}, \boldsymbol{v}|\boldsymbol{d}_{w,obs}, \boldsymbol{d}_{s,obs}) \approx c_0 f(\epsilon_{global} = 0|\boldsymbol{v})f(\boldsymbol{v}|\boldsymbol{b})f(\boldsymbol{d}_w = \boldsymbol{d}_{w,obs}|\boldsymbol{b})f(\boldsymbol{b}). \tag{12}$$

The above equation was obtained by substituting our definitions for seismic and well data likelihoods into the last expression of equation 5. The ease of sampling according to $f(\boldsymbol{d}_w = \boldsymbol{d}_{w,obs}|\boldsymbol{b})$ will be dependent on the dimensionality of $\boldsymbol{d}_w$. While this might be tractable for well data of moderate dimensionality, generating samples according to $f(\epsilon_{global} = 0|\boldsymbol{v})$, i.e., models with null RMO errors in expectation over all subsurface locations, will typically be impractical to attain due to following factors:

1. Implicit parameterization of the velocity model through global parameters $\boldsymbol{b}$. As a result, finding a velocity model which optimally focusses the data in expectation at all subsurface locations might be challenging.

2. Dissociating the basin modeling and seismic imaging components of the joint inverse problem makes it challenging to leverage information from seismic to guide sampling of the prior basin models. As highlighted previously, solving equation 4 directly will be computationally prohibitive given the computational demands of the various forward modeling routines involved.

Our proposed solution to these challenges is to employ approximate Bayesian computation (ABC; Beaumont, 2010; Blum, 2010). Mathematically, we are assuming the following approximation for the posterior distribution $f(\boldsymbol{b}, \boldsymbol{v}|\boldsymbol{d}_{w,obs}, \boldsymbol{d}_{s,obs})$

$$\approx c_1 f(\epsilon_{global} < \tau_2|\boldsymbol{v}) f(\boldsymbol{v}|\boldsymbol{b}) f\big(g(\parallel \boldsymbol{d}_w - \boldsymbol{d}_{w,obs} \parallel) < \tau_1|\boldsymbol{b}\big) f(\boldsymbol{b}) \qquad (13)$$

Here, $c_1$ is the normalization constant, $\parallel . \parallel$ is a distance measure and $\tau_1$ and $\tau_2$ are threshold values for well data mismatch and $\epsilon_{global}$ respectively. In other words, instead of exactly matching $\boldsymbol{d}_{w,obs}$ and $\epsilon_{global}$ (equation 12), fitting data within certain thresholds is taken to be the criteria for generating approximate samples of the posterior distribution. In order to sample from the approximate distribution shown in equation 13, we generate a number of samples from $f(\boldsymbol{b})$ such that a functional value for the well data mismatch $g(\parallel \boldsymbol{d}_w - \boldsymbol{d}_{w,obs} \parallel)$ falls below the threshold $\tau_1$. We assign $g(.)$ to be the likelihood distribution $f(\boldsymbol{d}_w = \boldsymbol{d}_{w,obs}|\boldsymbol{b})$, which is a function of the data mismatch as shown in equation 8. For each produced sample, rock physics modeling is employed to sample $f(\boldsymbol{v}|\boldsymbol{b})$. From the set of velocity samples, models satisfying $\epsilon_{global} < \tau_2$ will be retained.

Ideally, the thresholds will have to be set to very low values in order for the posterior approximation to hold reasonably. A practical approach to specifying the thresholds is to retain a small percentage of the prior samples, which exhibit the best fit to data, as approximate samples of the posterior (Beaumont et al., 2002). For instance, we could generate several thousand basin models and retain the maximum likelihood model from every set of 1000 models as a sample from $f(g(\parallel \boldsymbol{d}_w - \boldsymbol{d}_{w,obs} \parallel) < \tau_1|\boldsymbol{b})f(\boldsymbol{b})$. However, this is a naïve sampling strategy and will require a very large number of basin modeling simulations. We employ importance sampling (Liu, 2004) to alleviate this issue. Instead of sampling $f(\boldsymbol{b})$, we sample a proposal distribution $f_{proposal}(\boldsymbol{b})$ for the prior, such that samples from this distribution have greater probability of matching well data. This allows us to set a higher value for $\tau_1$, boosting the efficiency of the approach. The specific nature of $f_{proposal}(\boldsymbol{b})$ we employ is described in the 'Real Case application' section. During sampling according to $f(\epsilon_{global} < \tau_2|\boldsymbol{v})$, the appropriate number of models to be retained can be determined empirically by performing a visual analysis of images, as shown in the 'Real Case application' section.

The above procedure will generate samples from the proposal distribution for the posterior

$$c_2 f(\epsilon_{global} < \tau_2 | \boldsymbol{v}) f(\boldsymbol{v} | \boldsymbol{b}) f(g(\| \boldsymbol{d}_w - \boldsymbol{d}_{w,obs} \|) < \tau_1 | \boldsymbol{b}) f_{proposal}(\boldsymbol{b}), \quad (14)$$

where $c_2$ denotes the normalization constant. The bias introduced by sampling proposal distribution (equation 14) instead of the target distribution (equation 13), will be compensated by attaching an importance weight $w_j$ with every generated sample. For any sample, we assign weight $w_j$ as the ratio of target posterior density function to the proposal posterior density function

$$w_j = \frac{c_1 f(\boldsymbol{b}_j)}{c_2 f_{proposal}(\boldsymbol{b}_j)}. \quad (15)$$

Here, subscript $j$ denotes the sample index. For deriving the above result, we assumed that $f(\epsilon_{global} < \tau_2 | \boldsymbol{v})$ and $f(g(\| \boldsymbol{d}_w - \boldsymbol{d}_{w,obs} \|) < \tau_1 | \boldsymbol{b})$ in equations 13 and 14 evaluate to approximately equal values. Normalization constants, $c_1$ and $c_2$, are not known. Consequently, an additional normalization step is necessitated

$$w_j^{norm} = \frac{w_j}{\sum_{j=1}^{m} w_j}. \quad (16)$$

Normalized weight $w_j^{norm}$ facilitates specification of the distributions up to a normalization constant (Liu, 2004) and will serve as our importance weights. These weights will be used during evaluation of kernel estimates or statistics such as mean and variance of the target approximate posterior density.

**Pore pressure prediction and uncertainty analysis**

In addition to generating geologically consistent velocity models, another contribution of the proposed methodology is that it allows us to perform seismic calibration of basin modeling outputs. We will discuss these ideas with respect to pore pressure prediction (PPP). In basin modeling, pore pressure evolution in the basin is simulated by solving differential equations for sediment compaction mechanisms and fluid flow with boundary conditions designed to simulate physics of the problem. This enables us to model effects of common overpressure generating mechanisms such as compaction disequilibrium, aquathermal pressuring, smectite-illite transformation or kerogen maturation. However, usage of basin modeling pore pressure predictions to applications such as drilling hazard mitigation is generally limited because of lack of spatial data constraints in basin modeling. Velocity models are good indicators of overpressured formations and seismic velocity models are frequently employed for PPP. As discussed by Dutta (2002), this is primarily because many characteristics of overpressured rocks such as high porosities, low bulk densities, low effective stresses and high temperatures are strongly coupled to rock velocity. Constraining basin models with seismic velocity information will thus lead to greater reliability on the predicted pore pressures.

In our methodology, the dependency between pore pressure predictions from the basin model and velocity is established through the rock physics model. Note that overpressured formations will exhibit low effective stresses, leading to lower velocities. This is typical of clastics. The approximate inference scheme based on migration velocity analysis yields posterior velocity models honoring the kinematic information of the seismic data. Basin modeling pore pressure realizations implicitly linked to the sampled posterior velocity realizations will be used to perform uncertainty quantification for pore pressure. Estimated pore pressure uncertainty will conform to the prior geological uncertainty on any geopressuring mechanism modeled in the basin model, porosity, temperature and mudweight data collected at the well, as well as the posterior seismic velocity model.

## REAL CASE APPLICATION

In this section, we present application of our methodology in a north-central Gulf of Mexico basin located off the coast of Louisiana. Available dataset in this basin consists of log data from several wells, biostratigraphic data and P-Z summed 3D seismic data acquired at four-millisecond sampling using ocean bottom cables (OBC). The area where seismic data was recorded has shallow water depth, approximately 36 meters on average. The seismic source lines are perpendicular to the receiver lines. Source line spacing is 400 meters and source spacing is 50 meters, while receiver line spacing is 600 meters and receiver spacing is 50 meters. Maximum offset is about 6 kilometers. A legacy depth migrated seismic volume was also available. Ray-based tomographic inversion was used to obtain the isotropic velocity model for depth migration.

We present our analysis on the 2D section shown in Figure 2. A well is present along the section at which porosity data, obtained from bulk density log, mudweight data and bottom-hole temperature data is available for basin modeling calibration. Note that no repeat formation test (RFT) data were available, hence we used mudweight data as a guide to upper bound pore pressure. Gamma ray log and bio stratigraphic data available at the well were interpreted to identify major lithologic changes and key events on the geologic timescale. The cross-section has a spatial extent of about 37 km. Along the X-direction, the model was discretized into 600 grid points, with grid spacing of approximately 60 meters. We will model 10 depositional layers to the Upper Miocene. Corresponding horizons were interpreted on the seismic image to derive the structural model for the basin shown in <u>Figure 2</u>.

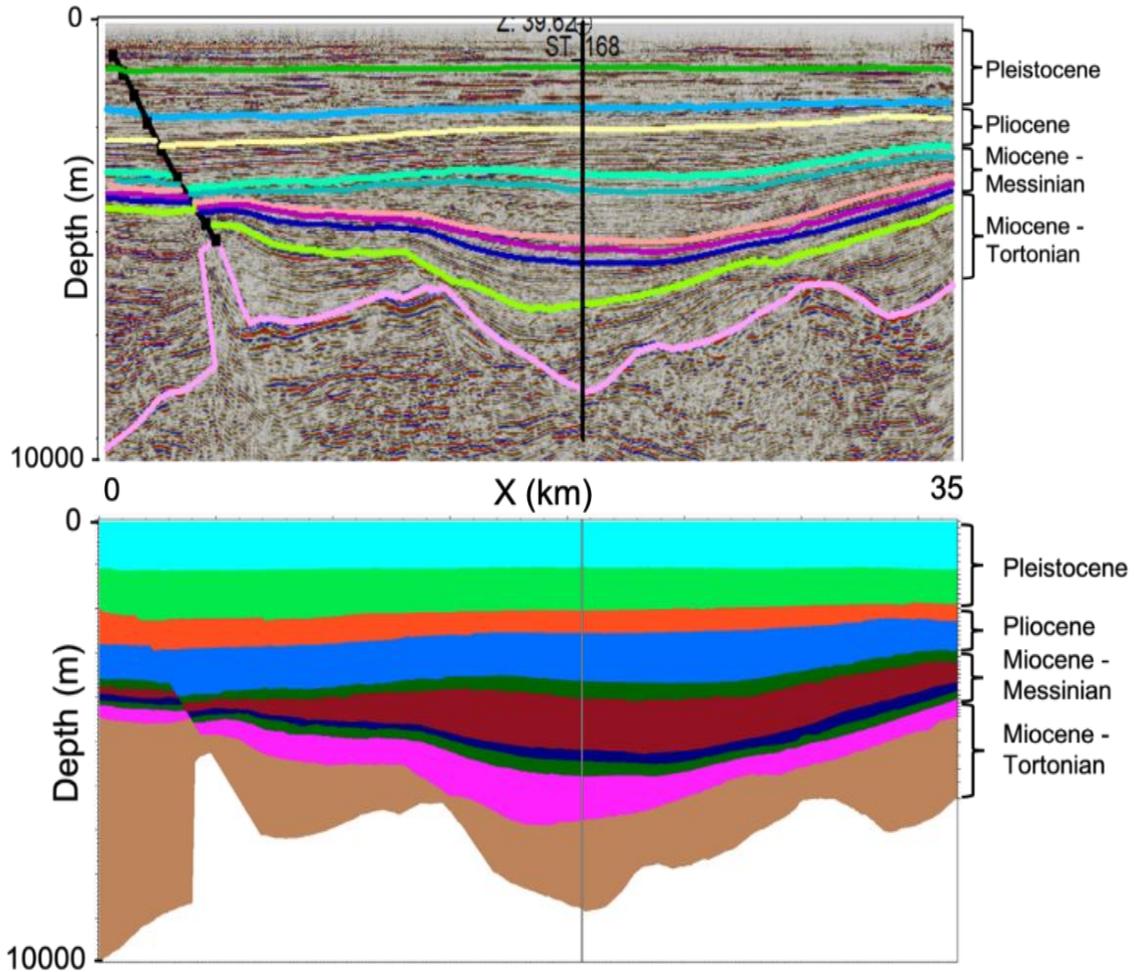

Figure 2: (Top) The depth migrated seismic section for which we present our analysis and interpretations of key geologic horizons (Bottom) Corresponding structural model of the basin

Along the depth direction, each lithologic interval was discretized into 10 equally spaced grid points between its top and base horizon. A fault is present in the cross-section, which was assumed to be a closed fault in the simulations. The seismic section also indicates presence of salt below the Upper Miocene layer. Our analysis is limited only to the supra-salt layers and we do not consider any effects from salt movement history in the basin. This is left as a future improvement of the model. Geological processes of sedimentation of these layers, mechanical compaction, fluid flow and heat flow in the basin were simulated from Upper Miocene till present day using the commercial software PetroMod. Mass balance equations, in conjunction with Darcy's law and sediment compaction laws, are used to simulate mechanical compaction and fluid flow. Heat flow is simulated using energy balance equations accounting for conductive, convective and radiogenic transport of crustal heat flow. Hantschel and Kauerauf (2009) provide a detailed treatment of the various physical transport laws used in PetroMod.

**Prior uncertainty and Monte-Carlo basin modeling**

The first component of our workflow constitutes defining the prior uncertainty on basin modeling parameters and performing Monte-Carlo basin simulations. Among the various sources of basin modeling uncertainties listed in the previous section, we consider the following: 1) Mechanical compaction or porosity reduction of the sediments captured by porosity-depth constitutive relations for each lithology 2) Fluid flow related uncertainty modeled by permeability-porosity relations 3) Heat flow boundary condition of the basal heat flow for the basin. Our choice was motivated by the fact these factors significantly control the variability of the present-day porosity, pore-pressure and mineral volume fractions which serve as inputs for rock physics modeling.

For each lithology, the porosity reduction law is specified according to an Athy type (Athy, 1930) law

$$\varphi = \varphi_0 e^{-c z_e}. \tag{17}$$

Here, $\varphi$ denotes porosity, $\varphi_0$ is the initial depositional porosity of the sediments, $c$ is Athy's compaction coefficient and $z_e$ denotes equivalent hydrostatic depth i.e., depth of the rock with same porosity and lithology under hydrostatic conditions. We underscore that the compaction law specified in equation 17 is formulated in terms of equivalent hydrostatic depth instead of the absolute depth. As noted by Hantschel and Kauerauf (2009), such a formulation can be analytically or numerically converted to two other commonly used porosity reduction laws formulated as porosity-effective stress or porosity-rock frame compressibility relations. The permeability-porosity constitutive relationship is specified according to the Kozeny-Carman type relation (Ungerer et al., 1990)

$$k_v = \begin{cases} F\frac{20\varphi'^5}{S^2(1-\varphi')^2} \ if \ \varphi' < 10\% \\ F\frac{0.2\varphi'^3}{S^2(1-\varphi')^2} \ if \ \varphi' > 10\% \end{cases}. \qquad (18)$$

Here, $k_v$ denotes vertical permeability, $F$ is a scaling factor, $S$ denotes specific surface area and $\varphi' = \varphi - 3.1 \times 10^{-10}S$. The horizontal permeability $k_h$ is expressed as $a_k \times k_v$, where $a_k$ represents the permeability anisotropy factor.

For each lithology, we consider $\varphi_0$, $c$, $S$, $F$ and $a_k$ as uncertain parameters. We define truncated Gaussian distributions as the prior probability distributions over the range of possible values that these parameters can assume. $\mu$ and $\sigma$ of the prior distributions for these parameters are shown in Table 1. Typical parameter values for pure sand and shale litholgies are available in literature (Hantschel and Kauerauf, 2009). For most lithologies, the depth-averaged clay volume fraction $vclay$ (Table 1), derived using the gamma ray log, was used as a guide to assign $\mu$, by roughly interpolating between pure sand and shale values. Standard deviations $\sigma$ were assigned such as to keep a fairly large uncertainty. Gamma ray log coverage did not extend to the top and bottom lithologies and hence larger uncertainty was assumed for these lithologies. The upper and lower limits of the truncated Gaussian distributions were fixed for each parameter type as shown in [Table 2](#). Preliminary investigations revealed that, by assigning sand-like properties to lithology 6 and 8 with $vclay = 0.36$, we were unable to explain some of the high mudweight pressure values. Note that $vclay$ is a depth-averaged property and thus these lithologies might have thin shaly beds. A possible reason for the discrepancy observed with mudweights is that we were precluding possibilities of pressure buildup by assigning sand-like properties to the entire lithological section. Hence, prior distributions for these two lithologies were kept same as those with unknown lithologies. The basal heat flow was assumed to be spatially and temporally constant. Uncertainty of the heat flow value was specified as Gaussian distribution with mean and standard deviation of 47 and 3 mW/m$^2$ respectively.

The uncertain prior model space for the basin thus consists of 51 parameters in total: 5 lithological parameters for each of the 10 lithologies and the basal heat flow. Value assignment of any other input parameters to the basin model was carried out by standard techniques employed in typical basin modeling workflows (Hantschel and Kauerauf, 2009). For instance, the sediment water interface temperature was assigned according to the method proposed by Wygrala (1989), while thermal properties of the lithologies were assigned by mixing standard parameter values for sand and shale rocks in proportion to the lithological average clay volume fractions. Subsequent to specification of prior uncertainty and basin modeling inputs, Monte Carlo sampling was performed to generate 2500 prior samples of the uncertain parameters. 2500 basin modeling runs were performed to simulate the porosity, pore-pressure and temperature evolution from Upper Miocene to present day. [Figure 3](#) shows the present-day porosity, pore pressure, and temperature sections for two prior realizations. We assumed that during deposition, clay minerals in the shale sediments consist of 80% smectite and 20 % illite components. As discussed by Dutta (1987, 2016), this is a reasonable assumption for Gulf Coast shales. We follow Dutta's approach for tying the chemical diagenesis of clay minerals resulting in smectite-illite transition with the thermal history of the basin. Specifically, we use first order Arrhenius rate theory for modeling the chemical reaction controlling smectite-illite transition. After illitization, the clay minerals consist of 20% smectite and 80% illite. Cross sections of illite volume fraction in the clay mineral calculated following this approach are shown in Figure 3.

Table 1: Prior distributions $\mathcal{N}(\mu, \sigma)$ of the uncertain lithological parameters

| Lithology | Vclay | $c$ (km$^{-1}$) | $\varphi_0$ | $S$ ($10^7$ m$^{-1}$) | $F$ | $a_k$ |
|-----------|-------|-----------------|-------------|------------------------|-----|-------|
| Lithology 1 | Unknown | $\mathcal{N}(0.5, 0.2)$ | $\mathcal{N}(0.55, 0.06)$ | $\mathcal{N}(5.05, 4)$ | $\mathcal{N}(2.45, 6)$ | $\mathcal{N}(3.10, 1.5)$ |
| Lithology 2 | 0.60 | $\mathcal{N}(0.5, 0.18)$ | $\mathcal{N}(0.56, 0.04)$ | $\mathcal{N}(7.04, 3)$ | $\mathcal{N}(1.48, 4)$ | $\mathcal{N}(2.22, 1.2)$ |
| Lithology 3 | 0.55 | $\mathcal{N}(0.44, 0.18)$ | $\mathcal{N}(0.54, 0.04)$ | $\mathcal{N}(6.54, 3)$ | $\mathcal{N}(1.91, 4)$ | $\mathcal{N}(2.41, 1.2)$ |
| Lithology 4 | 0.45 | $\mathcal{N}(0.34, 0.18)$ | $\mathcal{N}(0.51, 0.04)$ | $\mathcal{N}(5.0, 3)$ | $\mathcal{N}(3.15, 4)$ | $\mathcal{N}(3.29, 1.2)$ |
| Lithology 5 | 0.50 | $\mathcal{N}(0.39, 0.18)$ | $\mathcal{N}(0.53, 0.04)$ | $\mathcal{N}(6.05, 3)$ | $\mathcal{N}(2.50, 4)$ | $\mathcal{N}(2.60, 1.2)$ |
| Lithology 6 | 0.36 | $\mathcal{N}(0.5, 0.2)$ | $\mathcal{N}(0.55, 0.06)$ | $\mathcal{N}(5.05, 4)$ | $\mathcal{N}(2.45, 6)$ | $\mathcal{N}(3.10, 1.5)$ |
| Lithology 7 | 0.70 | $\mathcal{N}(0.55, 0.18)$ | $\mathcal{N}(0.60, 0.04)$ | $\mathcal{N}(8.03, 3)$ | $\mathcal{N}(0.89, 4)$ | $\mathcal{N}(1.84, 1.2)$ |
| Lithology 8 | 0.36 | $\mathcal{N}(0.5, 0.2)$ | $\mathcal{N}(0.55, 0.06)$ | $\mathcal{N}(5.05, 4)$ | $\mathcal{N}(2.45, 6)$ | $\mathcal{N}(3.10, 1.5)$ |
| Lithology 9 | 0.60 | $\mathcal{N}(0.5, 0.18)$ | $\mathcal{N}(0.56, 0.04)$ | $\mathcal{N}(7.04, 3)$ | $\mathcal{N}(1.48, 4)$ | $\mathcal{N}(2.22, 1.2)$ |
| Lithology 10 | Unknown | $\mathcal{N}(0.5, 0.2)$ | $\mathcal{N}(0.55, 0.06)$ | $\mathcal{N}(5.05, 4)$ | $\mathcal{N}(2.45, 6)$ | $\mathcal{N}(3.10, 1.5)$ |

Table 2: Upper and lower limits of the prior distributions shown in [Table 1]. The limits are fixed within each parameter type

| | $c$ (km$^{-1}$) | $\varphi_0$ | $S$ (10$^7$ m$^{-1}$) | $F$ | $a_k$ |
|---|---|---|---|---|---|
| **Lower limit** | 0.15 | 0.36 | 0.1 | 0.001 | 1.1 |
| **Upper limit** | 0.9 | 0.75 | 12 | 25 | 5.5 |

The next step is to determine the likelihood of the prior models conditional to the well calibration data. Calibration data consists of porosity, mudweight and bottom-hole temperature data (red circles in [Figure 4]). As discussed in the methodology section, we assume that each datatype is conditionally independent of the other given the basin modeling parameters $\boldsymbol{b}$. We also make the assumption that all data samples of any datatype are conditionally independent given $\boldsymbol{b}$. Thus, the likelihood for a particular datatype $f\left(\boldsymbol{d}_{w,obs}^{i} \mid \boldsymbol{b}\right)$ can be expressed as the product of the marginal likelihoods of all the data samples $\prod_{j=1}^{m} f\left(d_{w,obs}^{ij} \mid \boldsymbol{b}\right)$, where $d_{w,obs}^{ij}$ is the sample observed at a specific depth $j$ and $m$ is the total number of data samples. The marginal likelihood model for the porosity data samples is assumed to have a Gaussian distribution centered at the observed value, with a standard deviation of $10^{-2}$ to reflect the uncertainty associated with deriving the porosity data from bulk density log by making assumptions about the grain and fluid density. The marginal likelihood distribution for mudweight data constitutes a truncated triangular distribution with the observed mudweight and hydrostatic pressure as the upper and lower limits respectively. Note that mudweight refers to the density of the drilling fluid and is usually kept higher than the true formation pressure. The recorded mudweights can thus be taken to be the upper limits on the pore pressure. Hydrostatic pressure represents the theoretical lower limit on the pore pressure at any given depth. At any depth, if the driller keeps the mudweight to be significantly higher than the corresponding hydrostatic pressure, it will typically be indicative of an overpressured formation. We chose a truncated triangular distribution for pore pressure likelihood such that it assigns maximum probability to the recorded mudweight and minimum probability to the hydrostatic pressure. Probabilities for intermediate values between the two limits are linearly scaled. Bottom-hole temperature data typically underestimate/overestimate the true formation temperature and appropriate corrections need to be applied. The temperature data shown in Figure 4 were obtained after applying Horner correction to the data (Horner, 1951). We specified likelihood for the corrected bottom-hole temperature data points as Gaussian distributions centered at observed value, with a standard deviation of 2 $^0$C to account for uncertainty in the data and correction procedure. Figure 5 depicts these likelihood distributions. In Figure 4, we compare the present-day porosity, pore-pressure and temperature profiles of the 2500 prior models extracted at the well location against 30 posterior models that have the highest joint data likelihoods. It can be observed that the porosity outputs of these 30 models are tightly constrained in regions where calibration data is available and display high range of uncertainty in the absence of data as expected. Similarly, given the likelihood model for pore-pressure, models predicting pore pressures higher than the recorded mudweight are assigned a likelihood value of zero.

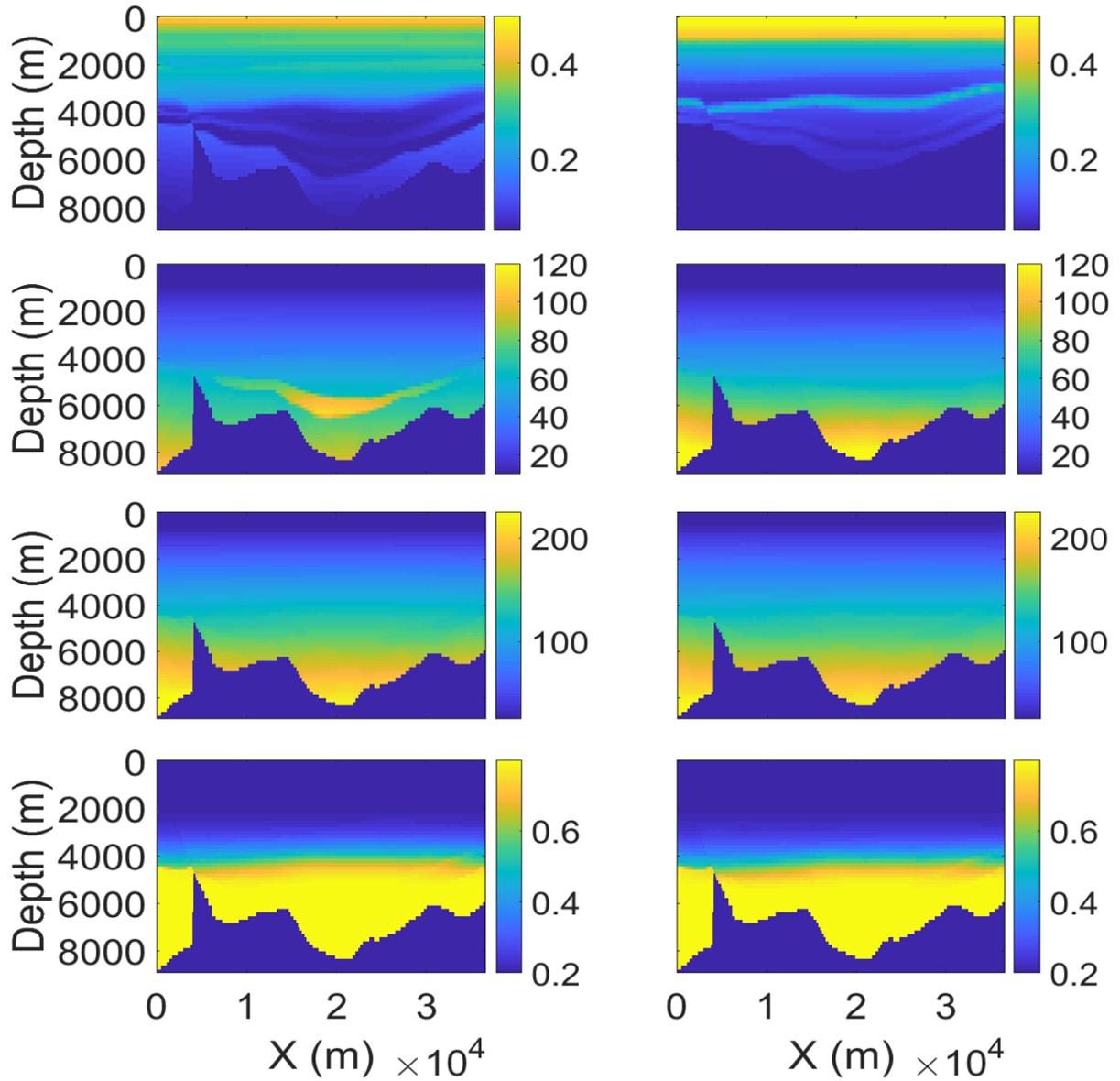

Figure 3: Present-day porosity (top row), pore pressure in MPa (second row), temperature in $^0$C (third row) and illite volume fraction in clay mineral (bottom row) sections of 2 randomly selected prior models (out of 2500), obtained after running the basin simulations

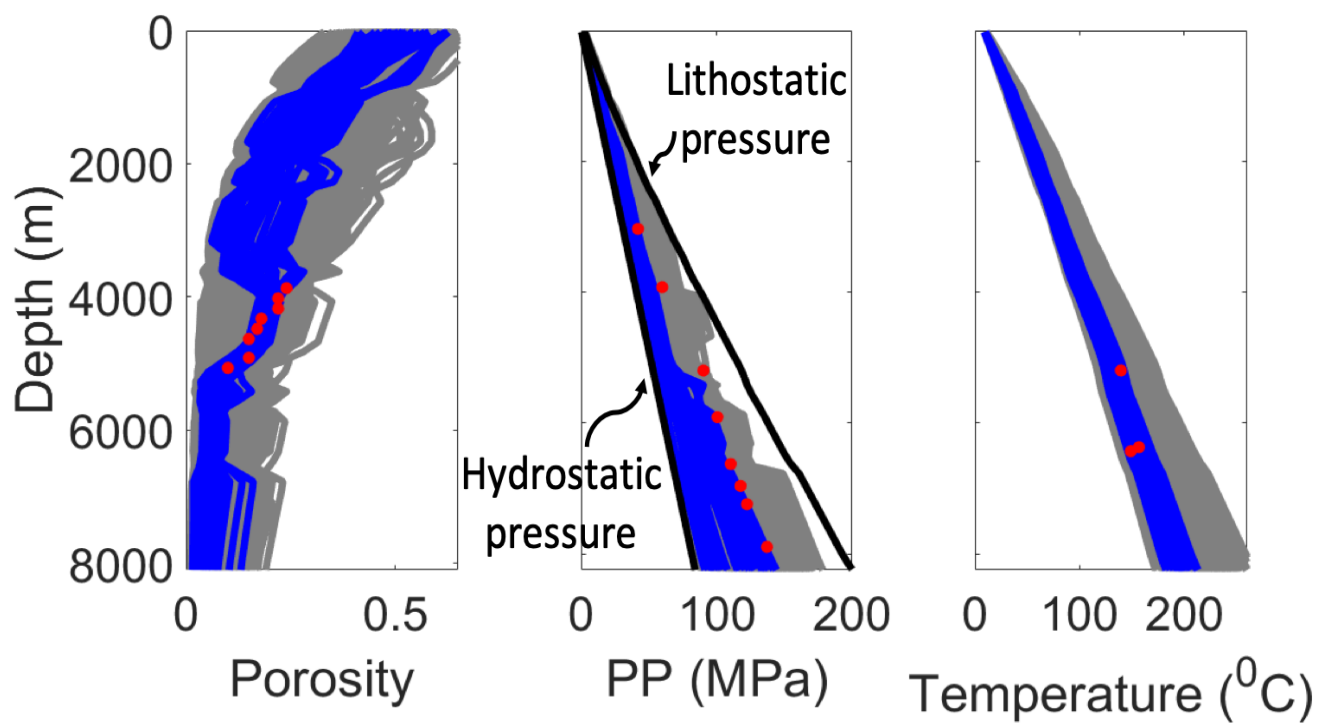

Figure 4: Porosity (left), pore pressure (middle) and temperature (right) profiles extracted at the well location from the corresponding present-day output sections of the 2500 basin simulations are shown in gray. Shown in blue are the outputs of 30 models with the highest likelihood. Observed calibration data are plotted as red circles.

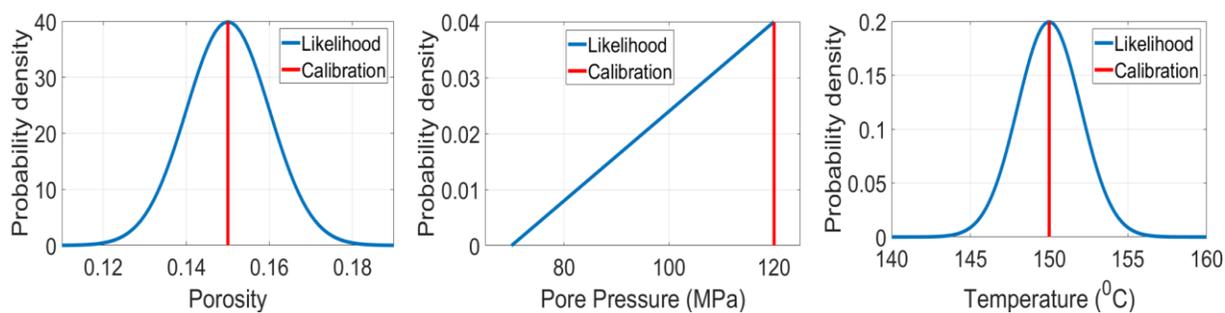

Figure 5: Marginal likelihood models for porosity (left) mudweight (middle) and bottom-hole temperature (right) calibration data.

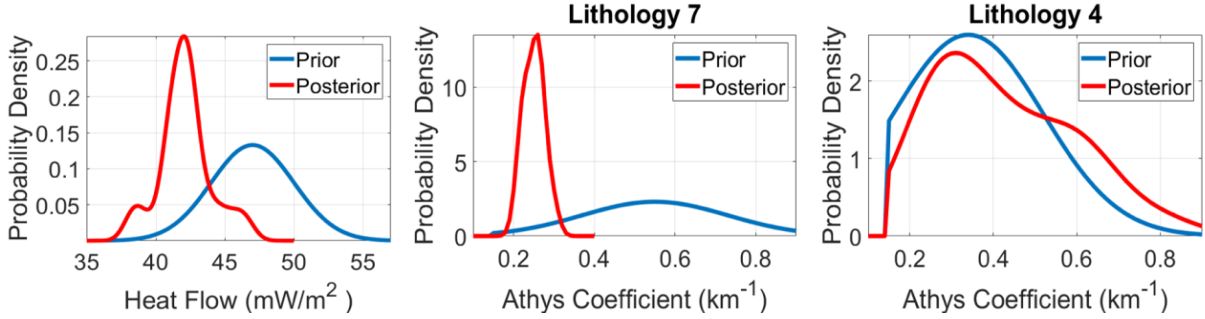

Figure 6: Comparison of prior and posterior distributions of basin modeling parameters after probabilistic calibration to well data

We define $f_{proposal}(\boldsymbol{b})$ as the product of the marginal posterior distributions

$$f_{proposal}(\boldsymbol{b}) = \prod_{i=1}^{51} f(b^i|\boldsymbol{d}_{w,obs}).$$ (19)

Note that the proposal distribution assumes independence between $\{b^i; i = 1, \ldots, 51\}$ even after conditioning to $\boldsymbol{d}_{w,obs}$, which is not necessarily a valid assumption. In order to account for this bias, we assign

$$w_j = \frac{c_1 \prod_{i=1}^{51} f(b_j^i)}{c_2 \prod_{i=1}^{51} f(b_j^i|\boldsymbol{d}_{w,obs})},$$ (20)

derived by appropriate substitutions in equation 15. In the above, $j = \{1, \ldots m\}$ and $m$ is the total number of samples generated. We generate 10000 samples of $\boldsymbol{b}$ from the proposal distribution and execute basin simulations. Subsequently, we evaluate model likelihoods $f(\boldsymbol{d}_{w,obs}|\boldsymbol{b})$ and select 1000 models with the highest likelihood values for rock physics modeling. Specifically, we are sampling according to the distribution $f(g(\| \boldsymbol{d}_w - \boldsymbol{d}_{w,obs} \|) < \tau_1|\boldsymbol{b})$ in equation 14. Since samples of $\boldsymbol{b}$ are obtained from the marginal posterior distributions, they will have higher chances of matching well data. Consequently, this allows setting a higher threshold value (10% of 10000) for ABC. Accepted 1000 models are referred as basin modeling posterior samples in subsequent treatment. [Figure 7](#) compares the well porosity profiles of the proposal models against the basin modeling posterior models.

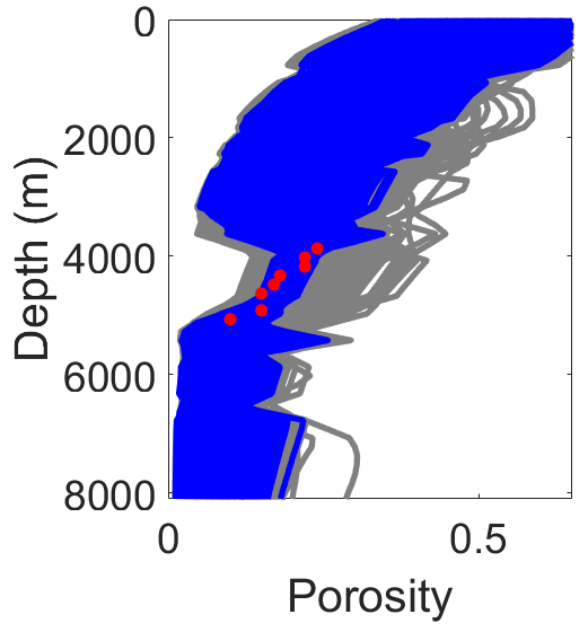

Figure 7: Present-day porosity profiles extracted at the well location from 10000 basin simulations obtained by sampling the proposal distribution for prior are shown in gray. Shown in blue are the outputs of 1000 models with the highest likelihood. Observed calibration data are plotted as red circles.

**Rock physics modeling and migration velocity analysis**

In this part, we sample $f(\boldsymbol{v}|\boldsymbol{b})$ using rock physics modeling. We underscore that we do not consider any modeling uncertainties and $f(\boldsymbol{v}|\boldsymbol{b})$ is deterministic. The rock physics model we use is the constant-clay model for shaly sands (Avseth et al., 2005). This model is analogous to the friable sand model proposed by Dvorkin and Nur (1996). We assumed four end members for the rock model: smectite, illite, non-clay stiff mineral composite for minerals such as quartz and feldspar, and brine saturated pores. Mineral end-member elastic properties ([Table 3](#)) were calibrated using the sonic log available at the well. During the calibration process, typical values for the elastic properties were randomly perturbed multiple times and rock physics modeling was performed for each case. Finally, values generating best fit with the sonic log were retained. Figure 8 compares the sonic log and velocity log modeled using the calibrated shaly sand model. The clay volume fraction is assumed to be constant within each lithology as listed in Table 1. The smectite-illite volume fractions are obtained using Dutta's (1987) approach as described earlier. Subsequently, the effective moduli of the solid mixed-mineral phase were calculated by the Voigt-Reuss-Hill average. This solid phase serves as the effective mineral end member for the shaly sand model. The other end member of the shaly sand model is the dry well-sorted solid phase at critical porosity, modeled as an elastic sphere pack subject to effective pressure using Hertz-Mindlin contact theory. The effective pressure is derived by Terzaghi's principle using the pore-pressure and lithostatic pressure outputs from basin modeling. Thus, the well sorted end member will vary for each accepted prior model depending on its effective pressure output. Given the end members of the shaly sand model, the elastic properties of the dry rock at any desired porosity is calculated using a modified lower Hashin-Shtrikman bound. Subsequently, effect of fluid substitution on the dry rock properties is determined using Gassmann's (Mavko et al., 2009) fluid substitution relations. Figure 9 shows 2 randomly selected velocity realizations obtained through the rock physics modeling.

Table 3: Calibrated mineral end member parameters

| Mineral | Parameter | Calibrated value |
|---|---|---|
| Non-clay stiff mineral | Bulk modulus | 34.43 GPa |
| | Shear modulus | 42.5 GPa |
| | Density | 2.63 g/cc |
| Smectite | Bulk modulus | 11.9 GPa |
| | Shear modulus | 3.2 GPa |
| | Density | 2.35 g/cc |
| Illite | Bulk modulus | 39.2 GPa |
| | Shear modulus | 17.3 GPa |
| | Density | 2.66 GPa |

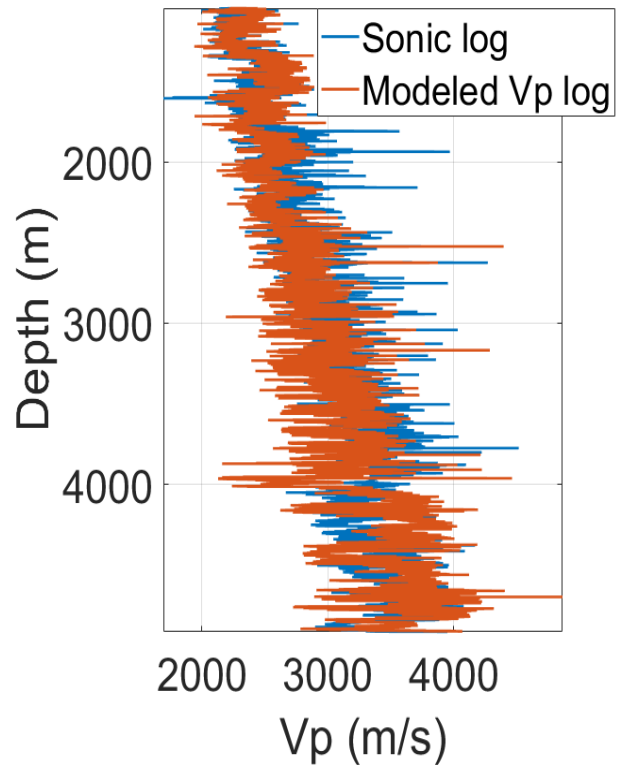

Figure 8: Sonic log and P-wave velocity (Vp) log obtained using the calibrated shaly sand model. Correlation coefficient is 80% between the true and modeled values.

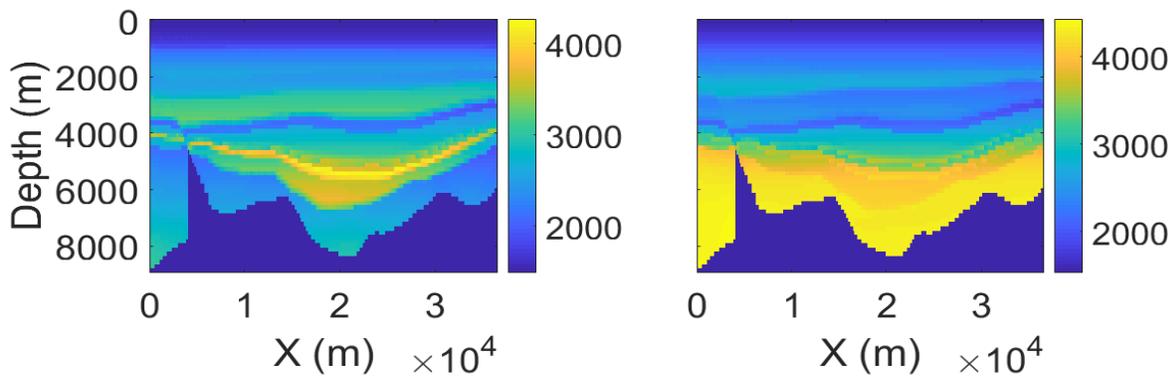

Figure 9: Two velocity (in m/s) realizations obtained through rock physics modeling using the outputs of the accepted basin models

In order to perform depth migration, we had to synthesize seismic data along the 2D cross-section from the original 3D data. This was necessary as simply extracting a 2D line, (either source line or receiver line), would result in under sampling and spatial aliasing due to the orthogonality of the source and receiver lines and sparse cross-line spacing. To synthesize 2D data, we chose a subset of the 3D data with midpoints within a one-kilometer swath, covering two receiver lines. The 2D cross-section of interest is located between these receiver lines. Figure 10 shows locations of the midpoints, two receiver lines, and corresponding sources. The positive x-indices refer to locations where the legacy velocity model and the depth image were provided to us. Assuming structures and velocity do not vary significantly in the cross-line direction, the chosen sources and receivers were rotated about their common midpoints (CMPs) to align in-line. The rotation is trusted to not change reflection moveouts because the water depth at this area is particularly shallow. As a result, no differential moveout correction was applied. Figure 11 shows locations of the sources and receivers after rotation. Rotated sources and receivers densely cover a patch one kilometer wide and 30 kilometers long, which we will regard as our 2D data. Note that the number of receivers increases in Figure 11 as compared to Figure 10. This is because every receiver records data from multiple shots and thus after rotation about CMP, it gets mapped to different locations for different shots. The resulting data were then sorted into 50-meter bins and stacked. This produces 536 shots spaced 200 meters apart along the 2D cross-section of interest. Figure 12 shows a sample shot gather. Some reflection hyperbolas are identifiable in the first four seconds of the data. Little coherent signal shows in the later parts of the data. For depth migration, we only use shots present in the positive x-locations shown in Figure 11 (approximately 20 kilometers long).

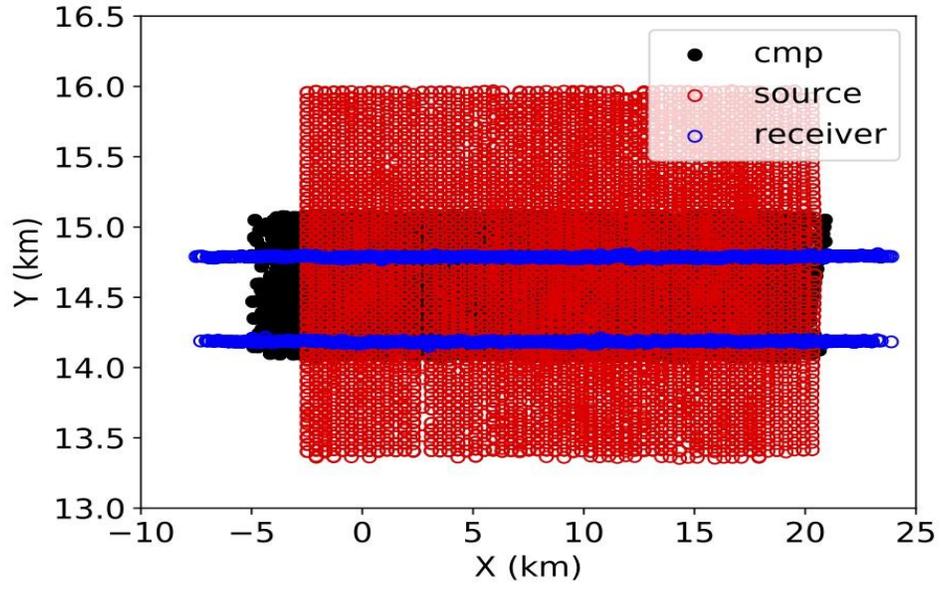

Figure 10: Locations of the two receiver lines surrounding the 2D cross-section of interest, corresponding sources and CMPs

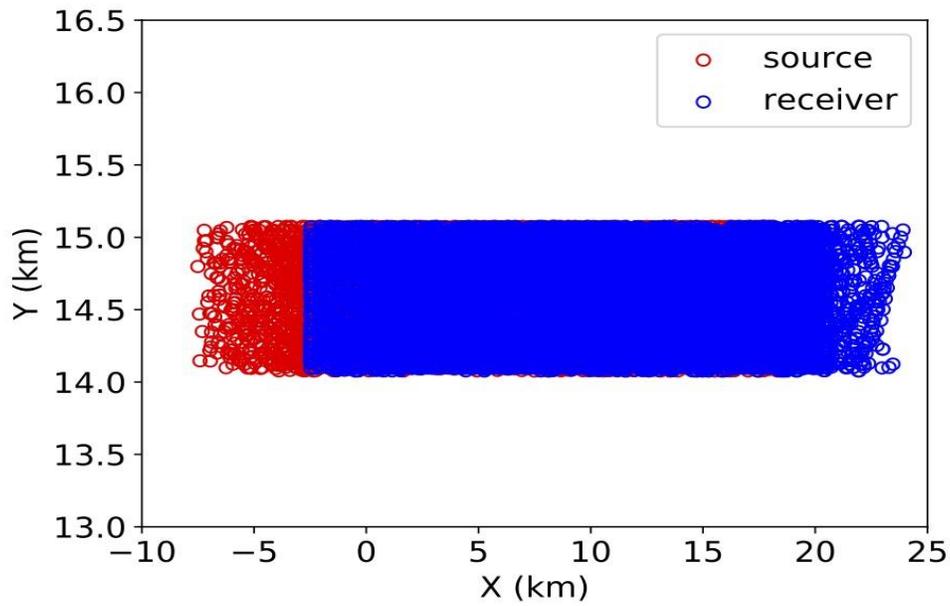

Figure 11: Locations of the sources and receivers after in-line rotation about CMPs for synthesis of 2D seismic data

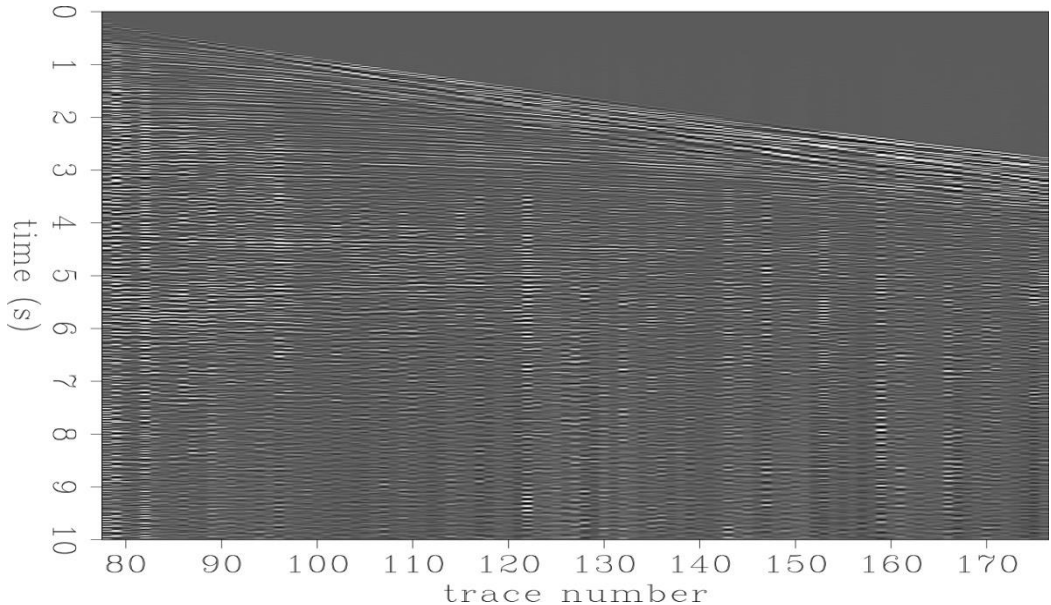

Figure 12: A shot gather generated after processing of seismic data along the 2D section

Reverse time migration (RTM; Alkhalifah, 2000) of the seismic data was performed with each basin modeling posterior velocity model to evaluate focusing quality of the imaged reflectors. RMO analysis was performed using ADCIGs in each case to estimate $\epsilon_{global}$ (equation [11]). In Figure 13, we plot $\epsilon_{global}$ values obtained with all the basin modeling posterior velocity models. For comparison, we also migrated the seismic data using the legacy velocity model estimated by ray-based tomographic inversion. It can be observed that the mean RMO error for a sizable number of models are distributed around the mean tomographic RMO error value, and some have lower mean RMO error. We ranked all the 1000 basin modeling posterior velocity models according to their $\epsilon_{global}$ values. In Figure 14, we show three velocity models with the highest ranks, i.e., lowest $\epsilon_{global}$ values, along with the legacy tomographic model. Corresponding migrated images are shown in Figure 15. It can be observed that most major reflectors present in the legacy image have also been focused with the basin modeling derived velocity models. In order to determine how many models need to be retained for the uncertainty analysis presented in the next sub-section, similar qualitative analyses of RTM images were conducted for the other top ranked models. We found that the top 30 models with lowest mean RMO errors generated images with acceptable focusing quality of the reflectors. The focusing quality of RTM images obtained with models falling above this threshold was found to deteriorate progressively in general (compare top against bottom row images in Figure 16). Note that $\epsilon_{global}$ is the RMO error averaged across all subsurface locations and consequently minimizing $\epsilon_{global}$ only ensures ADCIG flatness in a global sense. We use this measure to build the seismic data informed geological and physical model space for velocity, as shown in the next sub-section. If the goal is to derive velocity models with locally optimal ADCIG flatness, conventional velocity inversion formulations should be employed on this geological velocity model space to solve a local optimization problem for gather flatness.

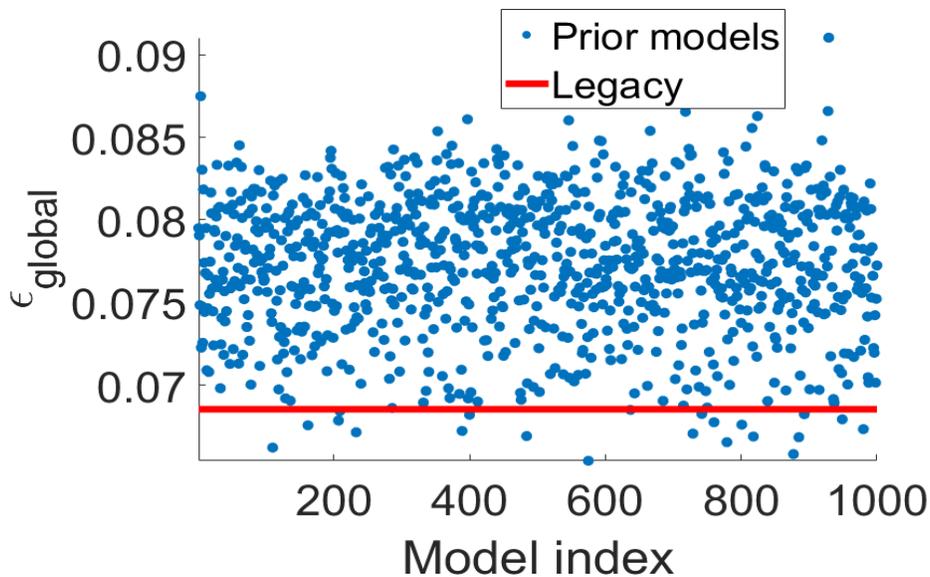

Figure 13: $\epsilon_{global}$ for 1000 basin modeling derived velocity models compared against legacy tomographic velocity model.

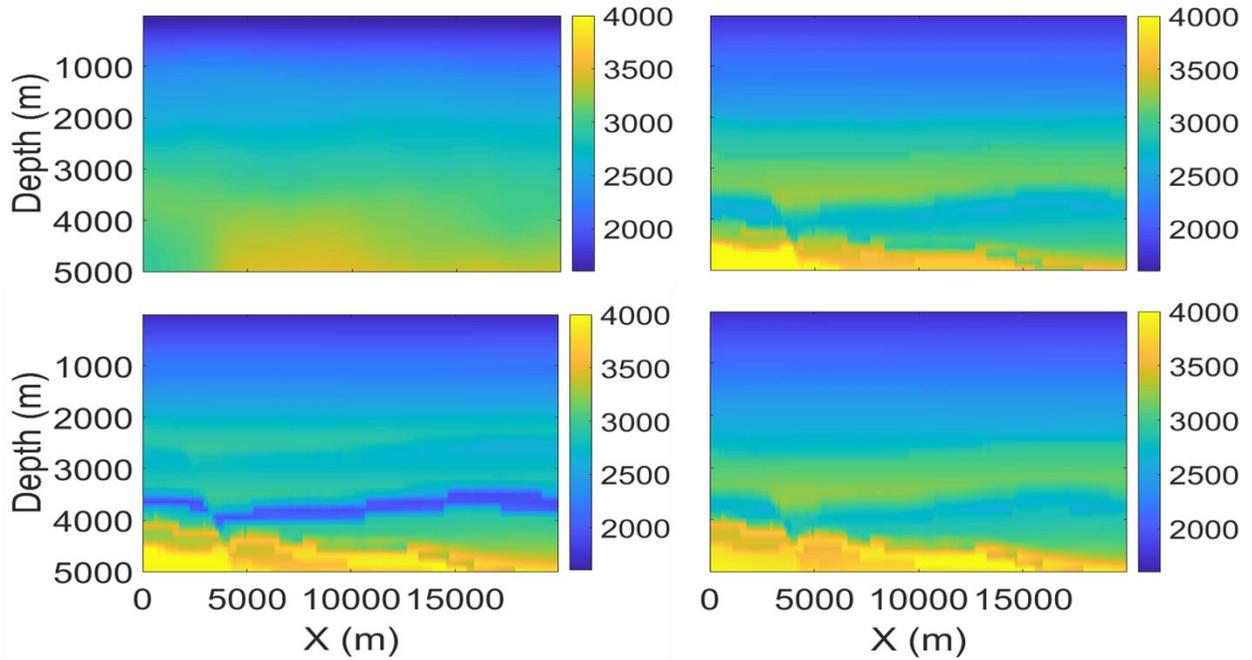

Figure 14: Legacy velocity (in m/s) model obtained tomographic inversion is shown in top left figure. Basin and rock physics modeling derived velocity sections with lowest $\epsilon_{global}$ values are shown in clockwise order from top right.

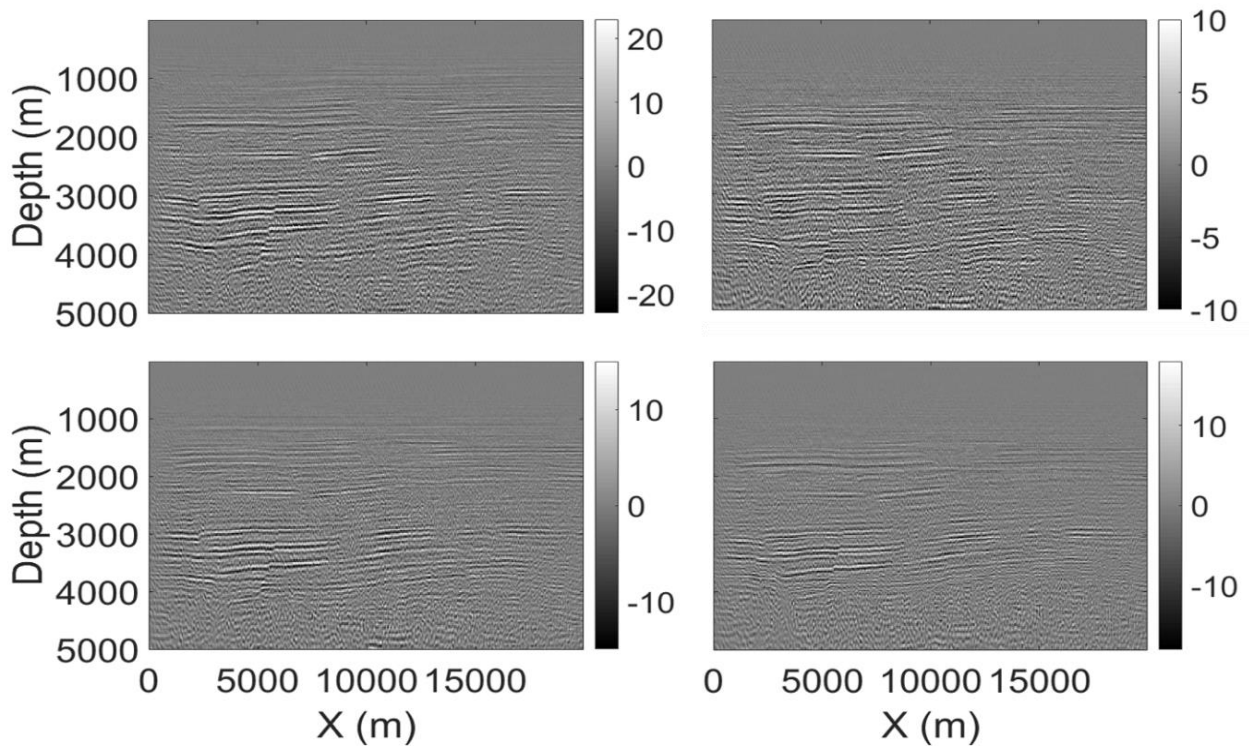

Figure 15: Reverse time migrated images obtained using corresponding velocity models shown in **Figure 14**.

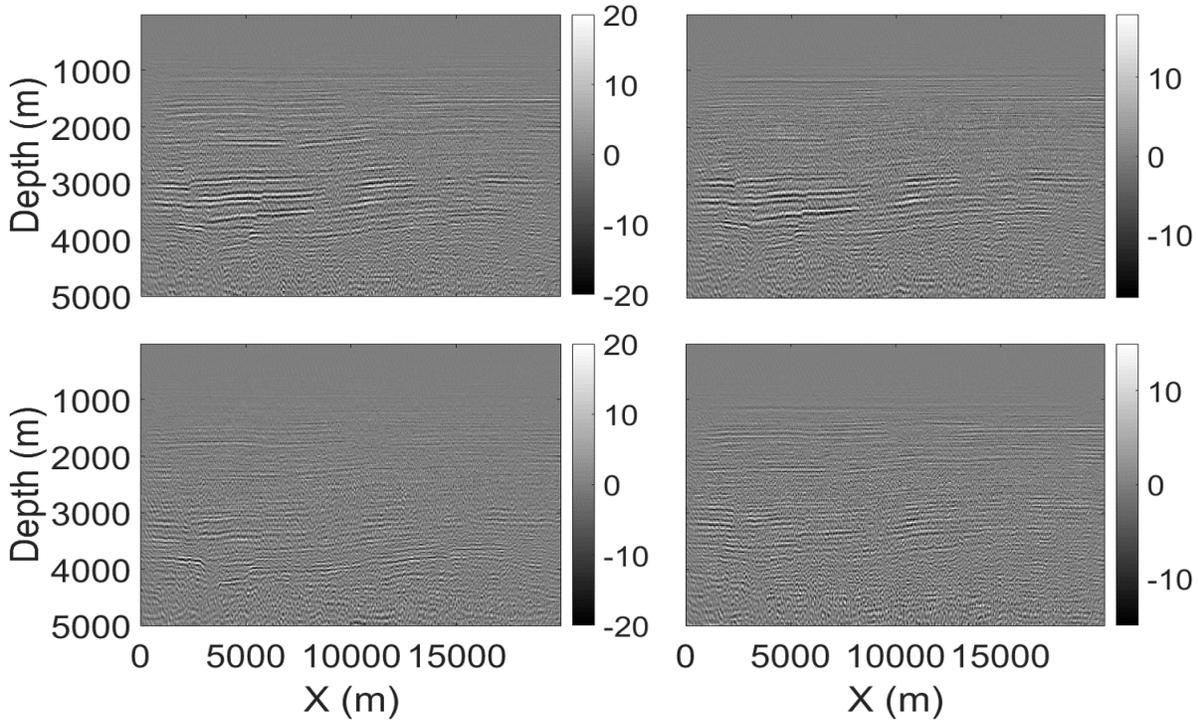

Figure 16: Reverse time migrated images for basin modeling derived velocity models ranked 18, 25, 38 and 50 (clockwise from top left) based on RMO analysis.

## Uncertainty quantification of velocity and pore pressure

For sampling according to $f(\epsilon_{global} < \tau_2 | \boldsymbol{v})$, we retain 30 models selected in the previous sub-section. For brevity, we term these accepted models as seismic posterior samples in subsequent discussions. In [Figure 17](#), we compare porosity and velocity depth profiles of the seismic posterior samples extracted at a particular surface location against the basin modeling posterior models. Porosity calibration data is available only in the Miocene layers and consequently basin modeling posterior realizations (shown in gray) have significant uncertainty in the Pleistocene and Pliocene layers. Incorporation of seismic kinematic information is effective in further constraining the uncertainty in these layers as evidenced by the spread of the seismic a posteriori realizations (shown in blue). Note that at a depth of about 4 km, the legacy tomographic velocity seems to be outside the range of geologically plausible velocities, even though it is an adequate velocity model for flattening the seismic gathers. Figure 18 shows the posterior velocity weighted mean and weighted standard deviation over the cross section estimated using the 30 accepted models. The weights used in the calculation of the mean and standard deviations are the normalized importance weights $w_j^{norm}$. Given our parameterization of velocity through basin models, we ensure that the variability in the velocity parameter space will conform to the geological history of the basin and associated uncertainty. This can be critical for velocity inverse problems such as full waveform inversion which are highly dependent on the quality of the starting velocity model for alleviating issues such as convergence to local minima.

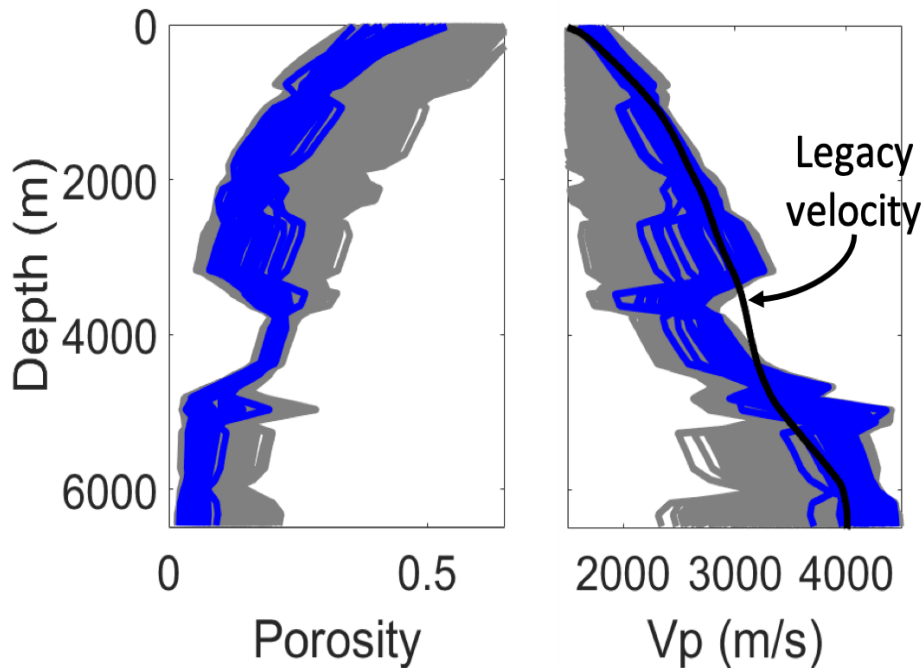

Figure 17: Porosity and velocity depth profiles extracted at surface x location of 15 km for the 1000 basin modeling posterior models derived. Overlain in blue are the profiles extracted from the 30 seismic posterior models derived after constraining to seismic kinematic information. Legacy velocity model trace at this location is shown in black.

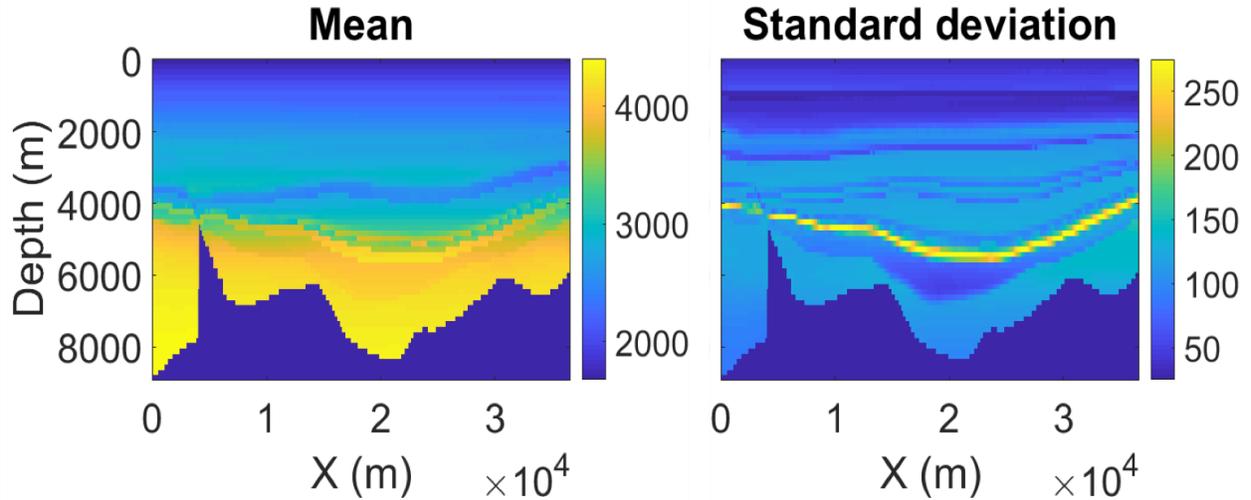

Figure 18: Posterior mean and standard deviation for velocity (in m/s) estimated using the 30 models accepted after migration velocity analysis.

Since each posterior velocity model is linked to pore pressure sections generated as outputs of basin simulations, it is also possible to update the pore pressure uncertainty according to kinematic information contained in seismic data. Figure 19 and Figure 20 depict how the prior pore pressure uncertainty was successively updated with information from mudweight and seismic data. We present the analysis with 2500 models obtained by Monte-Carlo sampling of the prior $f(\boldsymbol{b})$, 1000 models calibrated to well data only, and 30 models from the 1000 accepted after migration velocity analysis. In Figure 19, we compare the pore pressure depth profiles of these models extracted at surface location of 15 km. Figure 20 compares the prior and approximate posterior probability densities at various depths at this location. Note that $w_j^{norm}$ are also used during kernel estimation of the densities. It can be observed that seismic posterior distributions consistently have more probability density for higher pore pressure values as compared to the prior and basin modeling posterior distributions. Note that at deeper depths (see 4500 m and 6000 m plots), mudweight data does not rule out hydrostatic pressure as a possibility. However, information from seismic data assigns negligible probability to hydrostatic pressure and shifts the probability density to higher pore pressure values. This shifting of densities towards high pore pressures is correlated with depth as expected. The seismic a posteriori realizations can be used to constrain the spatial pore pressure uncertainty as shown in Figure 21 where we estimate the weighted mean pore pressure model and corresponding weighted standard deviation.

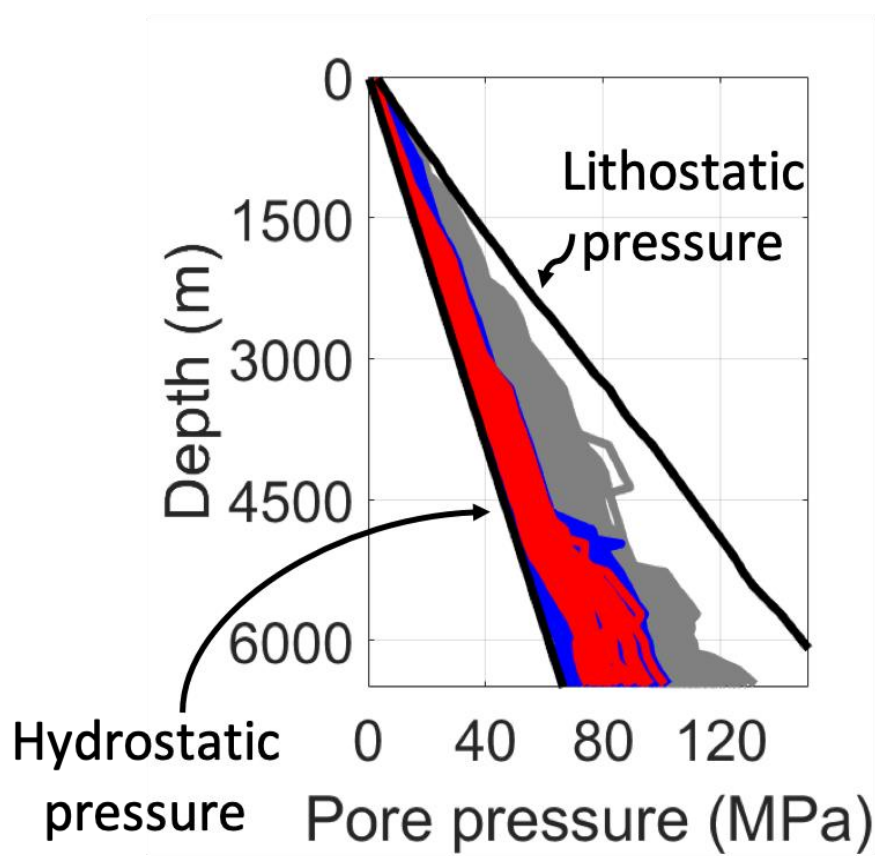

Figure 19: Pore pressure depth profiles extracted at surface x location of 15 km for 2500 prior models (shown in gray), 1000 basin modeling posterior models (shown in blue) and 30 seismic posterior models (shown in red).

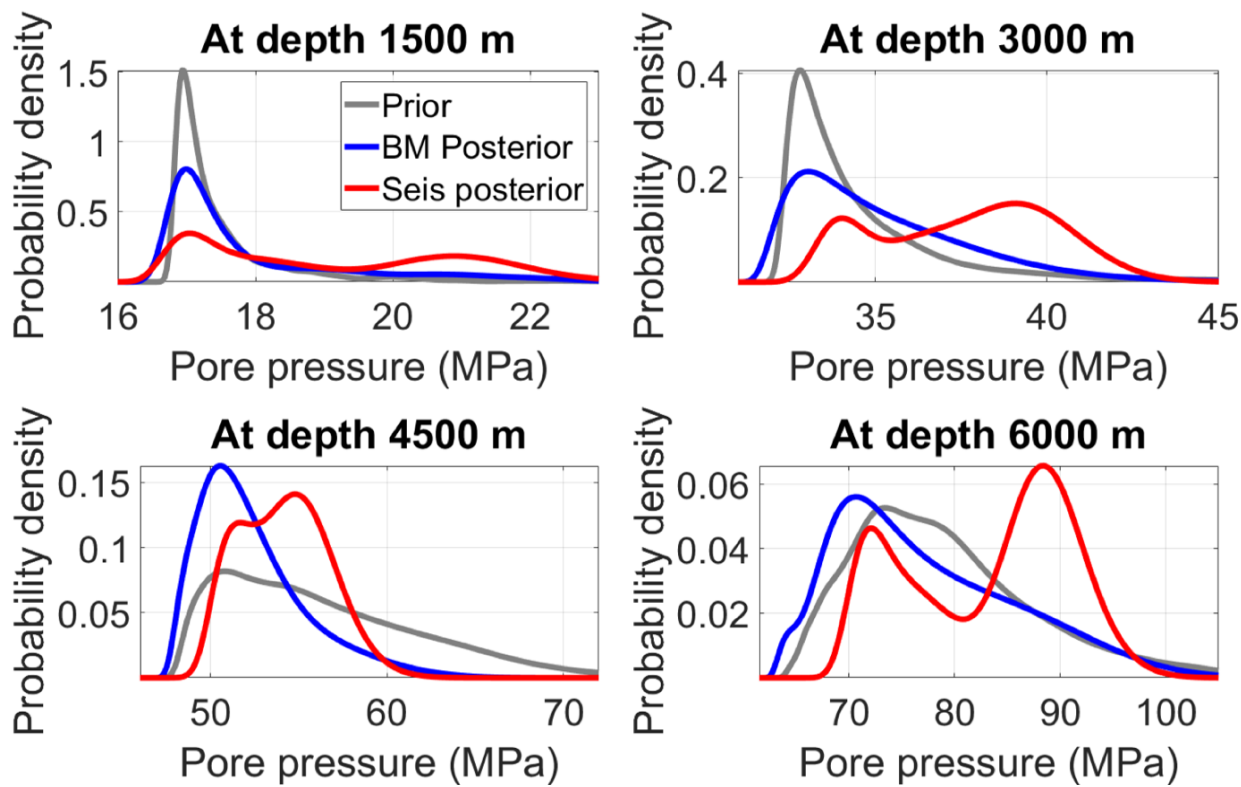

Figure 20: Estimates of pore pressure uncertainty estimated at x location of 15 km and at depths listed on top of each plot. Shown is the initial prior on the pore pressure (gray), the updated posterior distributions after constraining to well calibration data (blue) and further updated posterior after constraining to seismic kinematic information (red).

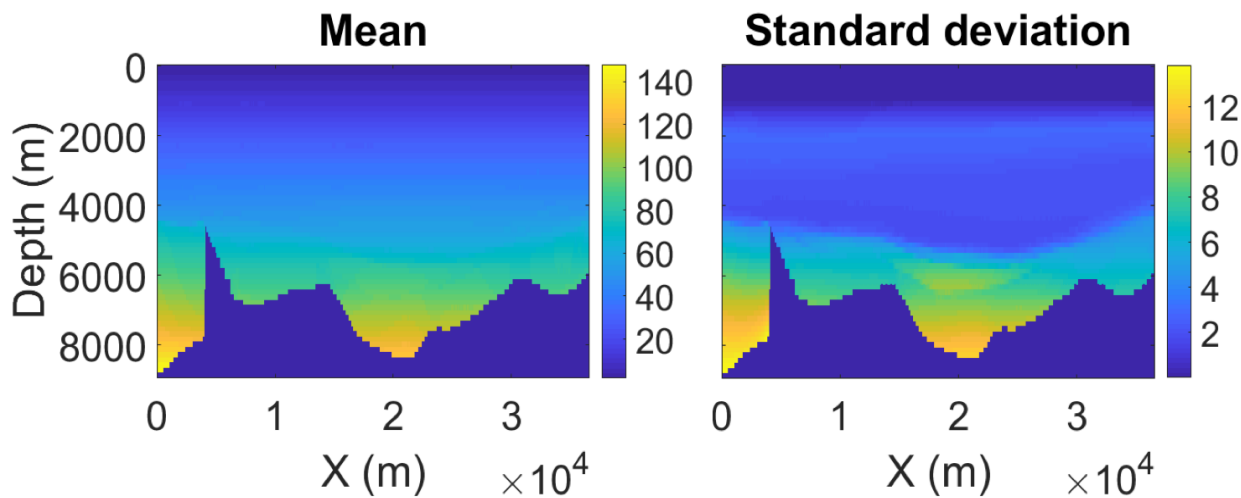

Figure 21: Posterior mean and standard deviation for pore pressure (in MPa) estimated using the 30 models accepted after migration velocity analysis.

DISCUSSION

In this section, we will discuss limitations of the presented methodology and potential solutions. A major practical challenge associated with the proposed approach is that it relies on performing multiple basin modeling, rock physics and depth migration evaluations, each of which can be computationally demanding. Considering additional sources of variability such as velocity anisotropy or increasing the model dimensionality could exacerbate the computational costs of this approach. We discussed several strategies for improving the feasibility of the approach such as decoupling the basin modeling and imaging components and employing strategies such as importance sampling. For 3D cases, it might be necessary to employ sampling methods such as Markov Chain Monte Carlo for sampling important regions of the prior parameter space with minimal number of forward model evaluations.

Another limitation of the proposed approach is associated with the implicit parameterization of velocity model through basin modeling parameters. Since basin modeling parameters are global properties for each geological layer, it becomes difficult to perform a local optimization of the velocity model using local information from seismic data. However, note that such global parameterization is parsimonious. For instance, in the case study we presented, the earth model grid containing 60000 grid cells was parameterized implicitly by 51 basin modeling parameters. This can be especially advantageous for the goals of this paper since the prior uncertainty space can be efficiently explored to build a geological and physical model space. Once the desired prior model space is generated, a constrained local optimization of the velocity model using tomography or full waveform inversion can be performed. The presented real case study has limitations such as: 1) we did not account for structural uncertainty in the interpreted geologic horizons and faults or spatial uncertainties in the basin modeling parameters, 2) we did not account for salt movement history in basin simulations and 2) we did not consider velocity anisotropy. We suggested some general research directions in the 'Methodology' section to address these shortcomings.

## CONCLUSIONS

Any earth modeling endeavor should satisfy four fundamental criteria: 1) The model should be geologically consistent, i.e. it should conform to our knowledge and beliefs about the geologic history and structure 2) Model variability should be constrained by any bounds derived from our physical understanding of the rocks 3) The model should agree with all available data from wells and geophysical experiments, and 4) Modeling uncertainty should be rigorously quantified. In this paper, we proposed to achieve all these objectives for velocity and pore pressure modeling in a single framework by integrating basin modeling, rock physics modeling and seismic imaging. Basin and rock physics modeling are critical for building a geological and physical parameter space for velocity. Posing the problem in a Bayesian framework facilitates consistent propagation of uncertainties associated with each modeling domain and data type. Solving a joint Bayesian inference problem with different types of forward models, well data and seismic data is computationally challenging and necessitates use of approximations and smart sampling strategies. Approximate Bayesian computation, in conjunction with importance sampling, can be utilized to gain computational tractability during conditioning of proposed workflow to well and seismic data. Migration velocity analysis can be integrated within the proposed framework to update prior geologic uncertainty on velocity and pore pressure based on seismic kinematic information.

## ACKNOWLEDGEMENTS

This work is supported by the sponsors of the Stanford Center for Earth Resources Forecasting (SCERF), the Stanford Basin and Petroleum System Modeling (BPSM), and the Stanford Exploration Project (SEP) consortia. We would also like to thank Steve Graham, the Dean of the School of Earth, Energy, and Environmental Sciences at Stanford University for funding. We thank Allegra Hosford Scheirer for her insights and guidance regarding the basin modeling components of the workflow. We also thank Schlumberger Limited for providing us the dataset for the real case study shown in this paper.